%% file: sample-manuscript.tex
\documentclass[manuscript,screen,nonacm]{acmart}
\AtBeginDocument{
  }

\definecolor{hidden-draw}{RGB}{20,68,106}
\definecolor{hidden-pink}{RGB}{255,245,247}
\usepackage{multirow}
\usepackage{graphicx}
\usepackage{ulem}
\usepackage{url}

\usepackage{xcolor}
\usepackage{tikz}
\usepackage{etoolbox}

\usepackage[edges]{forest}
\usepackage{adjustbox}

\begin{document}

\title{A Survey of Long-Document Retrieval in the PLM and LLM Era}
\author{Minghan Li}
\email{mhli@suda.edu.cn}
\affiliation{%
  \institution{School of Computer Science and Technology, Soochow University}
  \city{Suzhou}
  \state{Jiangsu}
  \country{China}
  \postcode{215031}
}

\author{Miyang Luo}
\email{myluo@stu.suda.edu.cn}
\affiliation{%
  \institution{School of Computer Science and Technology, Soochow University}
  \city{Suzhou}
  \state{Jiangsu}
  \country{China}
  \postcode{215031}
}

\author{Tianrui Lv}
\email{trlvtrlv@stu.suda.edu.cn}
\affiliation{%
  \institution{School of Computer Science and Technology, Soochow University}
  \city{Suzhou}
  \state{Jiangsu}
  \country{China}
  \postcode{215031}
}

\author{Yishuai Zhang}
\email{yszhang13@stu.suda.edu.cn}
\affiliation{%
  \institution{School of Computer Science and Technology, Soochow University}
  \city{Suzhou}
  \state{Jiangsu}
  \country{China}
  \postcode{215031}
}

\author{Siqi Zhao}
\email{sqzhaosqzhao@stu.suda.edu.cn}
\affiliation{%
  \institution{School of Computer Science and Technology, Soochow University}
  \city{Suzhou}
  \state{Jiangsu}
  \country{China}
  \postcode{215031}
}

\author{Ercong Nie}
\email{nie@cis.lmu.de}
\affiliation{%
  \institution{Center for Information and Language Processing (CIS), LMU Munich}
  \city{Munich}
  \country{Germany}
  \postcode{80538}
}
\affiliation{%
  \institution{Munich Center for Machine Learning (MCML)}
  \city{Munich}
  \country{Germany}
  \postcode{80538}
}

\author{Guodong Zhou}
\email{gdzhou@suda.edu.cn}
\affiliation{%
  \institution{School of Computer Science and Technology, Soochow University}
  \city{Suzhou}
  \state{Jiangsu}
  \country{China}
  \postcode{215031}
}

\renewcommand{\shortauthors}{Li et al.}

\begin{abstract}
The rapid growth of long-form documents presents a fundamental challenge to information retrieval (IR). Their length, dispersed evidence, and complex structures demand approaches that go beyond standard passage-level techniques. This survey provides a comprehensive survey of long-document retrieval (LDR) and organizes existing work into three major eras of development. The survey traces progress from classical lexical and early neural models to pre-trained language models (PLMs) and large language models (LLMs). It highlights core paradigms such as passage aggregation, hierarchical encoding, efficient attention, and LLM-driven re-ranking and retrieval. In addition to methods, the survey summarizes domain-specific applications, available evaluation resources, and the benchmarks that shape empirical study. It identifies open challenges related to efficiency trade-offs, evidence localization, multimodal alignment, and interpretability. The survey primarily focuses on methodological advances, while also considering practical applications in diverse domains. The principal conclusion is that future progress in LDR will depend on hybrid solutions that combine sub-linear indexing, structure-aware modeling, and LLM-based reasoning, as identified through the methods and challenges systematized in this survey.
\end{abstract}




\maketitle

\section{Introduction and Contribution}
\label{sec:intro}

From locating a single critical clause in a multi-hundred-page legal contract to synthesizing evidence from thousands of biomedical research papers, the ability to retrieve precise information from long-form documents has become a foundational challenge in the modern information landscape. The exponential growth of digital information has yielded corpora where such documents-scientific articles and patents, statutes and judicial opinions, financial and technical reports, clinical notes, books, and multimedia-rich web pages are the primary carriers of knowledge. These artifacts routinely span thousands to tens of thousands of tokens and exhibit rich internal structure. Retrieving evidence from such materials are foundational to web search, legal discovery, scientific knowledge mining, clinical decision support, and enterprise intelligence. 
Yet the very properties that make long documents valuable, such as distributed evidence, hierarchical organization, and inter-document linkage, also make them challenging for conventional information retrieval pipelines.

We study long-document retrieval : given a query $q$ and a corpus $\mathcal{D}$ of long documents, the system returns a ranking over documents or intra-document units , ideally with span-level rationales. In practice, indices are built at multiple granularities, and relevance must couple document-level utility with coverage of supportive segments. LDR departs from standard ad hoc retrieval along three axes: (i) evidence dispersion, where relevant signals are scattered across many segments and must be aggregated; (ii) hierarchical and cross-document structure, where headings, citations, hyperlinks, and version graphs condition relevance; and (iii) computational constraints, since encoder and interaction costs scale unfavorably with length unless architectures or pipelines are adapted. The problem further generalizes to long-query regimes and to multimodal documents.

Classical lexical methods are robust and efficient but degrade on long texts due to verbosity bias, topical drift, and an inability to model cross-segment dependencies. Early neural rankers improve matching at paragraph scale but remain bounded by input windows and quadratic attention, while naïve truncation or sliding-window heuristics sacrifice global coherence and induce ordering bias. These limitations have driven three waves of techniques that progressively address LDR-specific challenges. The PLM era extended pretrained transformer encoders to long inputs through three key innovations. First, passage-based "divide-and-conquer" strategies learned to aggregate scores from individual chunks, as seen in models like BERT-MaxP/SumP and PARADE. Second, hierarchical models were developed to pool information over a document's explicit structure, with examples including KeyB and IDCM. Third, sparse and efficient attention mechanisms, such as those in Longformer and BigBird, were introduced to mitigate the quadratic cost of processing long sequences.

Building on these foundations, the LLM era has introduced instruction-following models for two primary roles: as powerful re-rankers, exemplified by listwise prompting approaches like RankGPT, and through fine-tuning as end-to-end retrievers, such as the bi-encoder variants RepLLaMA and RankLLaMA. This new wave is also accompanied by system-level innovations for long-context efficiency, including sparse attention for LLMs, prompt compression, and KV-cache reuse.

Evaluation for LDR requires care. Label sparsity in web- and enterprise-scale testbeds, mixed relevance units , and graded judgments complicate standard metrics. Beyond document-level nDCG/MAP/MRR, segment-level adaptations and hierarchical recall are often necessary, especially under partial judgments. Specialized protocols address long-query QBD settings and structure- or layout-aware retrieval where section anchors and page elements govern relevance. Recent datasets also target cross-lingual and multimodal regimes, aligning assessments with realistic use.

The methodologies reviewed in this survey are directly motivated by applications in which document longness is an intrinsic challenge. Legal retrieval requires case-to-case matching, statute version alignment, and multi-source research over documents with rhetorical roles and dense citation/version graphs. Biomedical literature search and clinical decision support rely on full-text evidence localization across PubMed/PMC and EHRs, increasingly with multimodal signals. Web search must retrieve from long-form news features and technical white papers, often aligning text with images, tables, or embedded media. Scientific paper retrieval benefits from document-level representations informed by citation signals and LLM-guided concept indices, while cross-lingual LDR requires bridging languages without losing span-level provenance. Across these domains, effective systems preserve hierarchy, aggregate dispersed evidence, and provide auditable rationales.

\subsection*{Positioning Against Existing Surveys}
\label{subsec:positioning}
While comprehensive surveys exist for the broader fields of neural information retrieval~\cite{guo2020deep}, efficient Transformer architectures~\cite{tay2023efficient}, and the general application of large language models to information retrieval~\cite{zhu2023large}, these works typically address long contexts as one of many challenges rather than as the central focus. Consequently, they do not provide a unified, end-to-end treatment of the specific problems intrinsic to LDR, such as evidence dispersion across vast texts, hierarchical structure modeling, and the evolution of evaluation protocols tailored for long-form content.

To the best of our knowledge, this is the first survey to offer a holistic and consolidated view specifically on \textit{long-document retrieval}, tracing its progression across three distinct eras: from classical lexical models to the latest PLM and LLM-based paradigms. Our work distinguishes itself by not only systematizing the models but also providing integrated guidance on domain-specific applications, evaluation benchmarks, and a forward-looking research agenda, thus offering a complete reference for the field.

\subsection*{Contributions}
\label{subsec:contrib}
To provide a clear roadmap for researchers and practitioners and to catalyze future innovation, this survey makes the following key contributions:
\begin{itemize}
\item \textbf{Unified problem formulation.} We formalize LDR across document-, passage-, and layout-level targets under both short- and long-query regimes (including QBD), providing a common framework to connect previously fragmented research threads.
\item \textbf{Taxonomy across three eras.} We systematize methods from lexical and early neural baselines to PLM-based passage/hierarchical/sparse-attention models and LLM-based retrievers/rerankers, clarifying how each class combats segment dilution, preserves global coherence, and manages computational cost.
\item \textbf{Holistic relevance in the LLM era.} We identify the field’s paradigm shift from local window voting to instruction-aligned, global modeling with span-grounded justification, analyzing listwise vs. pairwise prompting, calibration of LLM judgments, and hallucination risks.
\item \textbf{Efficiency principles for long contexts.} We distill design patterns: sparse attention, hierarchical pooling, query-focused routing, prompt compression, KV-cache reuse, and relate them to first-stage recall, re-ranking, and reading/generation.
\item \textbf{Application blueprints.} We present end-to-end system frameworks for settings such as law, biomedicine, and academia, which combine structure-aware indexing, graph-informed expansion, long-context re-ranking, and span-grounded generation, tied to domain constraints.
\item \textbf{A forward-looking research agenda.} We articulate critical gaps in the field, including label sparsity, robust long-query handling, faithful multimodal retrieval, and efficiency-effectiveness trade-offs, outlining promising directions for future work.
\end{itemize}

\noindent Roadmap. The survey first reviews pre-PLM baselines, then covers PLM-based LDR and LLM-based retrievers/rerankers with efficiency innovations and multimodal extensions. We next provide a comparative analysis, summarize datasets and evaluation protocols, and detail application blueprints in law, biomedicine, web/news, scientific retrieval, and cross-lingual settings. We conclude with open challenges and a research agenda.

\begin{figure}[htbp]
    \centering
    \includegraphics[width=1.0\textwidth]{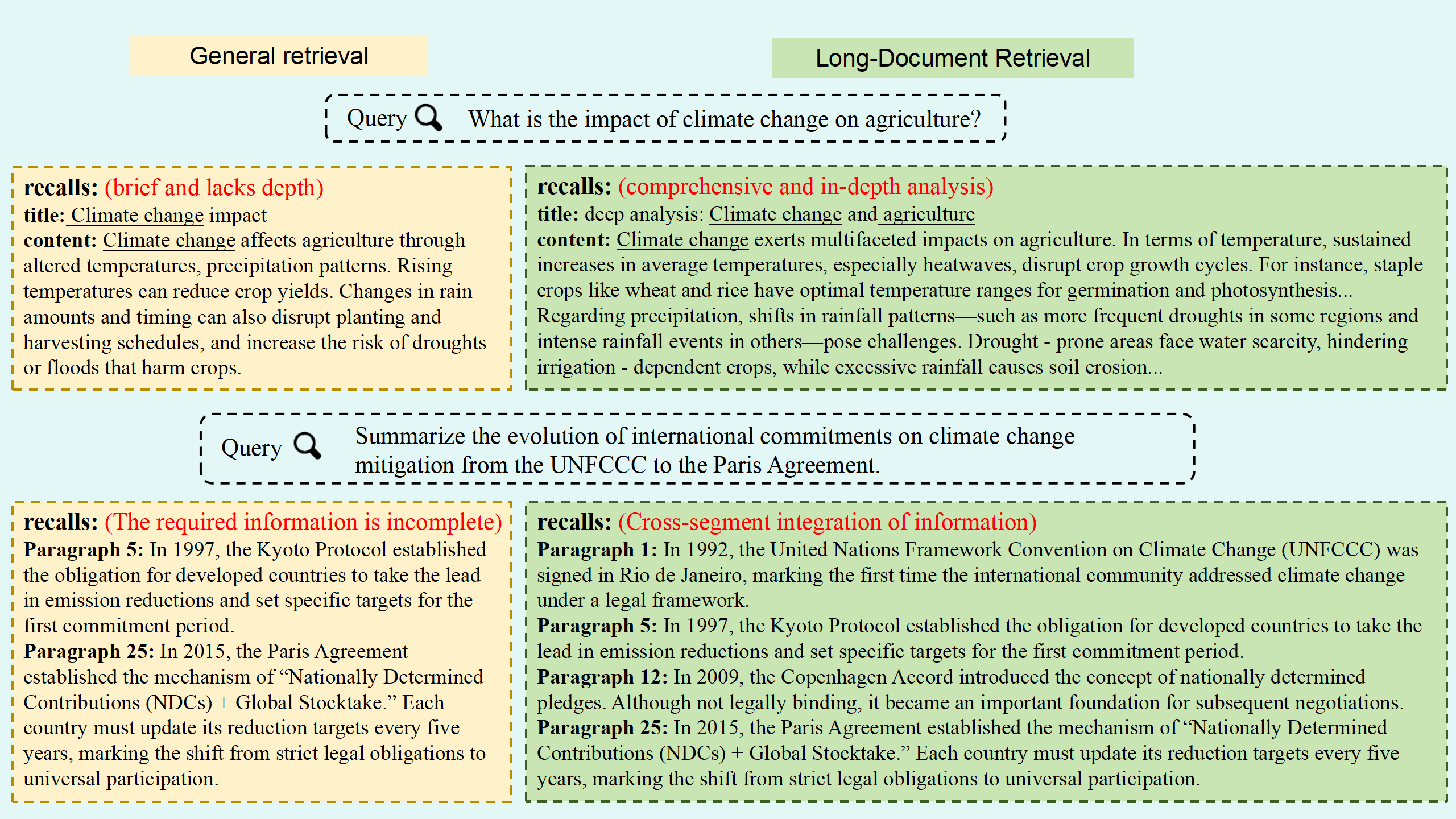} 
    \caption{In queries targeting long documents, a comparison between general retrieval methods and long-document retrieval methods reveals distinct differences: general retrieval methods struggle to acquire and present detailed, scattered information within long documents, whereas long-document retrieval methods excel at retrieving and organizing in-depth content from large volumes of textual resources.}
    \label{fig:retrieval_method_comparison}
\end{figure}

\input{new_taxonomy}

\section{Problem Scope And Definition}
Long-Document Retrieval is an important branch task in information retrieval. Its goal is to retrieve the most relevant documents or document fragments based on the user's query when facing a corpus containing a large number of long documents. Compared with classic document retrieval or paragraph retrieval tasks, LDR is unique in that it not only requires accurate identification of relevant information in a large document, but also requires efficient processing of redundant information and structural complexity while maintaining semantic integrity. As illustrated in Fig. \ref{fig:retrieval_method_comparison} , a clear distinction emerges between two types of retrieval methods when the query target is information within long documents. General retrieval methods struggle to capture scattered and detailed information completely, making it difficult to extract key content from long documents and present it effectively. In contrast, long-document retrieval methods are specifically designed to address this issue. Their core strength lies in the ability to accurately retrieve deep information from large-scale textual resources and organize such information in a structured manner, thereby fulfilling query requirements in long-document scenarios. The figure visually demonstrates the irreplaceability of long-document retrieval methods in specific contexts and provides clear guidance for the design of future methods. This capability gap also explains why "long documents" need to be defined from multiple dimensions (length, structure, and semantics). These characteristics are exactly the root causes of the ineffectiveness of general retrieval methods.

Currently, there is no unified standard for the definition of "long documents". Research usually defines it based on the following dimensions: (1) Length dimension: The document length far exceeds the maximum input limit of the standard Transformer model , often ranging from thousands to tens of thousands of tokens; (2) Structural dimension: Long documents usually have a clear hierarchical structure, such as chapters, paragraphs, titles, tables, and appendices, and the information distribution has obvious heterogeneity; (3) Semantic dimension: Long documents often contain multiple topics or multiple arguments, and there are problems of topic drift and long semantic span. Therefore, LDR not only faces the problem of information extraction, but also needs to pay attention to document structure perception and semantic focus capabilities. According to different retrieval objectives and downstream task requirements, long-document retrieval can be divided into the following task forms: (1) Document-level retrieval: using the entire document as the retrieval unit, returning the most relevant documents from the long document collection; (2) Passage/Section retrieval: using paragraphs, chapters or fragments generated by sliding windows as units, retrieving content fragments that are strongly relevant to the query; (3) Evidence retrieval: extracting fragments from documents that serve as the basis for answers for tasks such as question answering and reasoning, often used in multi-hop question answering or RAG systems; (4) Structure-aware retrieval: using the logical structure of the document for retrieval, such as title/paragraph tag/citation network, to improve semantic positioning accuracy; (5) Cross-document retrieval: supporting the aggregation and cross-comparison of relevant information in multiple long documents, targeting complex reasoning and content synthesis. Furthermore, LDR often intersects and integrates with other retrieval tasks, such as open-domain question answering, document reranking, summary generation, and retrieval-augmented generation (RAG), forming a complex information processing pipeline. Therefore, accurately defining the boundaries and forms of LDR is fundamental for subsequent modeling, evaluation, and method design.

\subsection{Core Challenges}

\subsubsection{Sparse and Inconsistent Supervision Signals}

In the majority of public benchmarks, relevance annotations for lengthy documents are typically concentrated at the document or paragraph level, while actual queries frequently involve only small fragments within the document. As a result of this discrepancy, the model's discriminative power is diminished, and fine-grained alignment becomes more difficult. For instance, 
the Natural Questions dataset introduced by Kwiatkowski\cite{kwiatkowski2019natural} provides paragraph-level long answers and brief answer annotations for each question, which somewhat parallel the query-local fragment correspondence. Nevertheless, this annotation type continues to fall short of addressing the fine-grained supervisory requirements at the sentence or even evidence chain level. Conversely, sparse relevance judgements are typically implemented by mainstream training and evaluation resources in the retrieval domain, including MS MARCO and TREC Deep Learning. The official MS MARCO leaderboard is based on pre-collected sparse judgements, as explicitly stated by Craswell\cite{craswell2020overview} in their TREC 2022 review. Additionally, the sparse character of MS MARCO labels is underscored in the TREC DL overview. Arabzadeh\cite{arabzadeh2022shallow} conducted an analysis of this sparse supervision method, observing that it may result in external validity issues in evaluations by obstructing the learning of more detailed matching patterns by models.

\subsubsection{Document Segmentation and Chunking}

To facilitate indexing and retrieval, it is frequently necessary to divide lengthy documents into smaller blocks, such as paragraphs or chunks. A long-standing challenge is determining the optimal segmentation without sacrificing semantic context. As a result of mechanical segmentation, traditional fixed-length chunking frequently disrupts sentence logic or conceptual units, resulting in "semantic fragmentation" during retrieval. Through multi-dataset analysis, Bhat\cite{bhat2025rethinking} systematically evaluated the influence of fixed-length chunking strategies on retrieval performance. They discovered that smaller chunks are more effective in short-answer scenarios, as they are able to accurately identify individual factual information. Conversely, larger chunks are more advantageous for tasks that necessitate a broader contextual understanding, such as thematic summarisation of long documents and novel plot analysis. Different embedding models exhibit varying sensitivities to block size, which they emphasise. For example, lightweight embedding models are more efficient when using smaller chunks, whereas large parameter models necessitate larger chunks to completely capture deep semantic information. This renders the selection of a segmentation strategy a complex yet essential step in the pursuit of a balance between semantic integrity and retrieval efficiency.

\subsubsection{Computational and Efficiency Issues}

The computational and storage overhead are substantially increased when long texts are processed. The retrieval throughput and latency are influenced by the larger index sizes and the increased number of word vectors or embeddings that must be computed when the documents are longer. This problem is especially evident in deep learning-based retrieval, particularly in Transformer-based re-ranking models. The computational complexity of the self-attention mechanism increases quadratically with the length of the input, rendering it nearly impossible to process ultra-long sequences in a single pass. Researchers have suggested a variety of enhancements in various directions to resolve this bottleneck. On the one hand, BigBird \cite{zaheer2020big} and Longformer \cite{beltagy2020longformer} employ sparse attention mechanisms to effectively enhance the model's capacity to manage lengthy sequences, thereby reducing computational complexity at the structural level. In contrast, Seo\cite{seo2024efficient} proposed Colour (Compression for Long Context Retrieval), an input-level approach that reduces input size by compressing paragraphs and retaining core information, thereby avoiding redundant computations caused by irrelevant content. The former enhances scalability by optimising the model structure, whereas the latter achieves cost reduction and efficiency gains through content compression.

\section{Pre-PLM Era}
\label{sec:preplm}
Before the advent of pre-trained Transformers, LDR relied on lexical statistics and early neural architectures. While these methods laid essential foundations, their scalability to book- or report-length inputs is limited by (i) sensitivity to topic drift and dispersed evidence, (ii) computational growth with document length, and (iii) loss of global discourse when truncation or coarse windowing is used. We summarize representative approaches and highlight how these models’ limitations motivate PLM/LLM-era designs.

\subsection{Lexical Methods: TF--IDF, BM25, and Heuristic Structure}
\label{subsec:lexical}

Classic vector-space retrieval \cite{JONES1973619} weights terms by within-document frequency (TF) and across-corpus rarity (IDF):
\begin{equation}
TF_{i,j} = \frac{n_{i,j}}{\sum_k n_{k,j}}, \qquad
IDF_i = \log \frac{|D|}{1 + |\{j : t_i \in d_j\}|}, \qquad
\mathrm{TF\text{-}IDF} = TF \times IDF.
\end{equation}

Where $n_{i,j}$ is the count of term $t_i$ in document $d_j$; $\sum_k n_{k,j}$ is the total number of terms in $d_j$; $TF_{i,j}$ (term frequency of $t_i$ in $d_j$) equals $n_{i,j}$ divided by $\sum_k n_{k,j}$; $|D|$ is the total number of documents in the corpus; $|\{j : t_i \in d_j\}|$ is the number of documents containing $t_i$ (the "$+1$" in IDF's denominator avoids division by zero); $IDF_i$ measures $t_i$'s rarity across the corpus; and $\mathrm{TF\text{-}IDF}$ is the product of $TF_{i,j}$ and $IDF_i$, quantifying $t_i$'s importance in $d_j$ relative to the corpus.

TF--IDF scales well and is interpretable, but ignores synonymy/polysemy and long-range semantics; in long documents, frequent background terms can dominate, and topic drift obscures sparse, query-relevant evidence.

Okapi BM25 \cite{robertson1995okapi} refines lexical matching with saturation and length normalization:
\begin{equation}
\label{eq:bm25}
\mathrm{score}(d,Q)=\sum_{q\in Q}\! \mathrm{IDF}(q)\cdot
\frac{\mathrm{TF}(q,d)\cdot (k_1+1)}{\mathrm{TF}(q,d)+k_1\!\cdot\!\bigl(1-b+b\cdot \tfrac{|d|}{\mathrm{avgdl}}\bigr)}.
\end{equation}

Where $\mathrm{score}(d,Q)$ is the relevance score between document $d$ and query $Q$; $d$ is the document evaluated; $Q$ represents the user's query; $q$ is an individual term in $Q$; $\mathrm{TF}(q,d)$ is $q$'s frequency in $d$; $k_1$ controls term frequency saturation; $b$ normalizes document length effects; $|d|$ is $d$'s term count; $\mathrm{avgdl}$ is the corpus average document length; and the score sums weighted contributions of each $q$, integrating $\mathrm{IDF}(q)$ (as in TF-IDF) and normalized term frequency adjusted by $k_1$ and $b$.

BM25 remains a strong, unsupervised baseline even for long documents due to robustness and efficiency \cite{10.1145/3534928}. Yet it is still lexical: dispersed, semantically related evidence without exact term overlap is easily missed.

Heuristic chunking and structure-aware weighting. To mitigate dispersion, classical systems adopt paragraph/passage indexing and structure-aware priors. Callan's segmented indexing \cite{10.1007/978-1-4471-2099-5_31} scores overlapping windows and then aggregates at the document level, reducing the chance that relevant text is split across boundaries. Structured metadata weighting \cite{Kowalski2002} emphasizes salient sections (e.g., abstract, conclusions, legal clauses). These heuristics help, but their fixed granularity and rule dependence struggle with heterogeneous long documents and cross-section reasoning.

Why lexical methods struggle with long documents. (1) \textit{Signal dilution}: relevance spans few sentences amid thousands of tokens; global bag-of-words mixing drowns weak signals. (2) \textit{Topic drift}: multi-topic, long narratives produce misleading TF/IDF statistics. (3) \textit{Structure unawareness}: discourse, section hierarchy, and cross-references are not modeled, limiting holistic understanding.

\subsection{Early Neural IR: Representation vs.\ Interaction Models}
\label{subsec:earlyneural}
Neural IR prior to Transformers pursued two lines: (i) representation-based models map queries/documents to vectors and compare them; (ii) interaction-based models compute fine-grained similarity matrices and learn matching patterns. Both improved beyond lexical matching, yet long-document scalability remained a core obstacle.

DSSM (representation-based) \cite{10.1145/2505515.2505665} projects queries and documents into a shared semantic space (cosine similarity), with letter-$n$gram hashing for vocabulary compression. While effective on web search logs, single-vector representations suffer semantic dilution on long, multi-topic documents, and lack explicit modeling of dispersed evidence.

DRMM (interaction-based) \cite{10.1145/2983323.2983769} builds histogram features of query-term $\!\times\!$ document-term similarities and aggregates with a term-gating network (e.g., IDF). Its focus on exact/strong matches suits ad-hoc retrieval, including longer pages; however, the implicit interaction matrix scales with $O(|q|\!\cdot\!|d|)$, making very long inputs costly without aggressive candidate pruning.

Treating matching as image recognition, MatchPyramid (interaction-based) \cite{10.5555/3016100.3016292} applies CNNs over the query–document similarity matrix with dynamic pooling to handle variable length. For long documents, the $|q|\!\times\!|d|$ map becomes large; dynamic pooling may discard subtle, far-apart evidence, and the model lacks explicit hierarchical aggregation.

PACRR (interaction-based) \cite{hui2017pacrr} convolves over the similarity matrix with $n$-gram kernels and uses $k$-max pooling per query term, followed by an LSTM combiner. To keep computation tractable, inputs are truncated or selected via \textit{firstk}/\textit{kwindow}, which reintroduces positional bias and risks missing late-occurring evidence in long documents.

DeepRank (query-centric contexts) \cite{pang2017deeprank} detects query-term occurrences and extracts local windows for focused matching, then aggregates across positions with positional decay and term importance. This avoids building a full similarity matrix and is robust to noisy long pages. Its main limitation is reliance on lexical overlap and fixed windows, which can miss semantically relevant, cross-sentence evidence spanning larger discourse units.

Why early neural models struggle with long documents. (1) \textit{Quadratic or linear-in-length interaction cost}: similarity maps and late-interaction pipelines scale with $|d|$, straining memory/latency on thousands of tokens. (2) \textit{Truncation/windowing side-effects}: necessary length control breaks discourse and induces head/tail bias; critical evidence after cutoffs is lost. (3) \textit{Insufficient global structure}: CNN/RNN over local patterns lacks explicit modeling of document hierarchy, cross-section dependencies, and global coherence demanded by long-form inputs.

Lexical and early neural methods introduced efficient baselines and fine-grained matching, but their assumptions (bag-of-words, fixed windows, global single vectors, or full similarity matrices) misalign with the length, structure, and dispersed evidence of long documents. These limitations directly motivate PLM-era strategies: efficient/sparse long-context attention, divide-and-conquer aggregation, and select-then-process cascades, as well as the LLM-era use of holistic rerankers with long-context optimizations.

\section{PLM and LLM Era Models for Long-Document Retrieval}

The advent of Pre-trained Language Models  and Large Language Models  has marked a paradigm shift in the field of long-document retrieval. To address the challenges posed by extensive contexts, researchers have moved beyond traditional bag-of-words models and shallow neural networks, developing a range of more powerful and sophisticated modeling strategies.

\begin{figure}[htbp]
    \centering
    \includegraphics[width=1.0\textwidth]{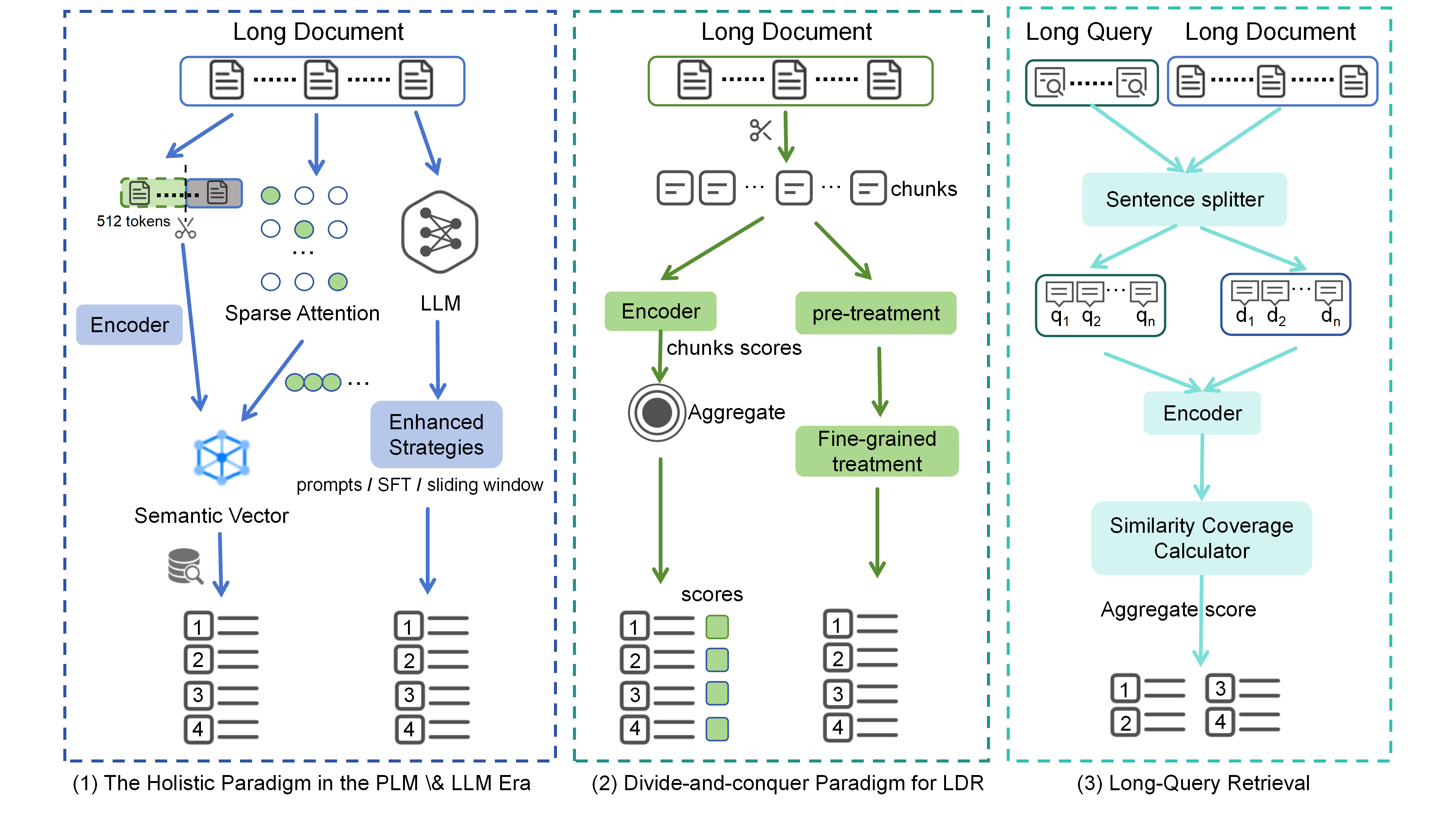} 
    \caption{An overview of the long-document retrieval paradigm in the PLM and LLM era. Methods evolve from (1) The Holistic Paradigm in the PLM \& LLM Era to (2) Divide-and-conquer Paradigm for Long Documents and (3) Long-Query Retrieval, reflecting the field’s progression in balancing effectiveness, efficiency, and scalability.}
    \label{fig:LDRSolutionOverView}
\end{figure}

To systematically organize these cutting-edge approaches, this chapter categorizes them into several core paradigms, whose architectural overviews are illustrated in \textbf{Fig. \ref{fig:LDRSolutionOverView}}:
(1) \textbf{the Holistic Paradigm} (detailed in Section \ref{sec:Holistic Paradigm methods}), which aims to model and interact with the entire document as a single, indivisible unit;
(2) \textbf{the Divide-and-Conquer Paradigm} (detailed in Section \ref{Divide-and-Conquer Paradigm methods}), which segments long documents into smaller pieces for localized processing before aggregating global information;
(3) \textbf{Long-Query Retrieval} (detailed in Section \ref{subsec:qbd}), which tackles scenarios where queries themselves are long documents (e.g., legal case retrieval, patent prior-art search). Additionally, we discuss the \textbf{Indexing-Structure-Oriented Paradigm} in Section \ref{Indexing-Structure-Oriented Paradigm methods}, this paradigm focuses on innovating how documents are chunked and indexed (note: its workflows are highly diverse and difficult to consolidate into a single generalized diagram, so it is elaborated separately).

Each subgraph visually summarizes the core workflow of its corresponding paradigm, providing intuition before we delve into technical details.
The following sections will delve into the representative models and technical evolution within each of these paradigms.

\subsection{The Holistic Paradigm in the PLM \& LLM Era}
\label{sec:Holistic Paradigm methods}

The holistic paradigm aims to model the entire document or interact with the global context of the query-document pair, thereby avoiding the segmentation and pooling biases of traditional methods. The advantage of this approach lies in its ability to preserve the global semantics and structure of the document. However, it faces significant challenges in terms of computational cost and architecture, particularly when dealing with ultra-long documents. Despite these challenges, it performs well in tasks involving short or smaller-scale documents. The next section will explore the "Divide-and-Conquer Paradigm," which addresses the computational efficiency issues of the holistic paradigm by processing documents in smaller chunks, while also achieving significant results in long-document retrieval.

\subsubsection{Naive Truncation Baselines}
Early dense retrieval models such as DPR \citep{karpukhin2020dense} and ANCE \citep{xiong2020approximate} highlighted the potential of pre-trained language models to surpass lexical retrieval by leveraging contrastive learning. However, they are constrained by BERT's 512-token limit, and thus apply a naive truncation strategy: retaining only the first 512 tokens of a long document, based on the common assumption that important summary information resides at the beginning. This design introduces three critical flaws: (i) information loss, as relevant evidence beyond the cutoff is discarded; (ii) destruction of discourse structure, since long-range dependencies are severed; and (iii) positional bias, as the beginning of a document is always over-emphasized. Consequently, While DPR and ANCE pioneered dense retrieval, their truncation assumptions clearly fail in the context of long document IR. Truncation not only results in systematic information loss but also severely impairs the model's ability to model discourse-level relevance. Furthermore, the performance ceiling of this approach is very low, making it difficult to serve as an effective baseline for long-document retrieval.

\subsubsection{Long-Sequence Transformer Architectures}
A major research thrust has therefore been to extend Transformer architectures to efficiently handle long sequences. \textbf{Longformer} \citep{beltagy2020longformer} introduced a hybrid attention mechanism: a sliding window for local context ($O(n \times w)$ complexity) combined with global attention tokens to capture long-range dependencies. It established a foundation for models such as \textbf{BigBird} \citep{zaheer2020bigbird}, which used blockwise sparse attention and random connections to achieve linear scaling, and demonstrated state-of-the-art performance on long-document QA and summarization. Sparse attention models, such as Longformer and BigBird, significantly expand the context they can handle. However, their scalability to millions of documents remains questionable, especially in real-world retrieval scenarios, where models must simultaneously process multi-document interactions, not just the long context within a single document. Furthermore, cross-paragraph and cross-section evidence aggregation remains fragile, and sparse connectivity patterns can miss key matches, leading to low recall. \textbf{Reformer} \citep{kitaevreformer} replaced quadratic attention with locality-sensitive hashing, while \textbf{FlashAttention} \citep{dao2022flashattention} restructured attention into an IO-aware exact algorithm, reducing memory latency and enabling efficient $>16$k context. More recently, \textbf{LongNet} \citep{ding2023longnet} scales to billion-token inputs using dilated attention without catastrophic forgetting of shorter contexts.

Zhu et al.  introduced \textbf{LONGEMBED}\cite{zhu2024longembedextendingembeddingmodels}, a benchmark for long-context retrieval, alongside methods to extend context length in embedding models. They proposed both training-free and training-based strategies for Absolute Position Encoding (APE) models, enabling context expansion from 512 to 4k through position embedding reuse, interpolation, or fine-tuning. For Rotary Position Encoding (RoPE) models, they leveraged relative position encoding via self-extrapolation and NTK-aware interpolation to extend E5-Mistral’s context to 32k without compromising short-context performance.

Beyond efficiency-driven architectures, IR-specific adaptations emerged. The \textbf{QDS-Transformer} \citep{jiang2020long} integrates IR axioms (locality, hierarchy, query matching) into a structured sparse attention pattern:
\[
A_{\mathrm{QDS}} = A_{\text{local}} \cup A_{\text{sent}} \cup A_{\text{query}} \cup A_{\mathrm{[CLS]}},
\]
where $A_{\text{local}}$ enforces windowed interactions, $A_{\text{sent}}$ links tokens to sentence-level [SOS] markers, and $A_{\text{query}}$ globally connects document words to query tokens. On TREC DL'19, QDS-Transformer achieved improvements over strong baselines on nDCG@10 while cutting inference latency in half. By embedding IR axioms into attention models, QDS achieves improvements on small-scale benchmarks, but its highly task-specific sparse structure limits generalization. Its design assumptions (such as sentence-level [SOS] tagging) may not hold true in multi-domain or cross-lingual long-document retrieval, and its complex sparse kernel implementation hinders widespread replication.

Inspired by the “small-world” phenomenon, \textbf{Socialformer}\cite{zhou2022socialformer} dynamically samples long-range links based on centrality and query-aware distance. It partitions token graphs into “circles” and propagates information via intra- and inter-circle Transformers, a method that significantly outperforms BigBird on TREC DL tasks. Compared to static sparse patterns, Socialformer thus exemplifies a shift toward more data-adaptive and dynamic connectivity. A key practical advantage is its straightforward implementation using standard PyTorch libraries, which avoids the engineering overhead of custom CUDA kernels common in other sparse-attention models. However, its sampling-based strategy can introduce result instability, and the model's scalability beyond 8,000 tokens remains unverified.

M3-Embedding\cite{bge-m3}, as an innovative embedding model, offers solutions that cater to multilingualism, multifunctionality, and multigranularity. This model is capable of processing long documents containing up to 8,192 tokens and supports various functionalities, including dense retrieval, sparse retrieval, and multi-vector retrieval. M3-Embedding employs a novel self-knowledge distillation method that enhances training quality and retrieval accuracy by integrating relevance scores from different retrieval functions. Its efficient training strategy and robust multilingual support make it highly effective in multilingual retrieval tasks and long-document processing. Additionally, M3-Embedding optimizes its batch processing strategy, allowing for efficient handling of large-scale data, especially excelling in long-document retrieval and cross-lingual tasks.

\subsubsection{LLMs as Holistic Rerankers}
With the advent of large language models , holistic modeling has advanced from efficient Transformers to end-to-end ranking. \textbf{RankGPT} \citep{sun2023chatgpt} demonstrated that GPT-4 can act as a universal reranker by generating explicit ranked lists (e.g., ``[2]>[3]>[1]’’), rather than independent scores. Using a sliding-window strategy, it successfully reranked hundreds of passages, and permutation distillation further transferred this ability to a lightweight 440M DeBERTa, achieving superior nDCG@10 on BEIR with only 1k labels. RankGPT demonstrates the unique capabilities of LLM for listwise ranking. However, its reliance on sliding windows to join long documents can easily introduce query drift: the model may make inconsistent judgments for the same query in different windows. Furthermore, its high inference cost and prompt sensitivity make it difficult to deploy at scale \cite{hagstrom2025language}.

More systematically, \textbf{RepLLaMA} and \textbf{RankLLaMA} \citep{ma2024fine} explored fine-tuning LLaMA models as dense retrievers and rerankers, directly encoding full 2k-token documents without segmentation. RepLLaMA uses the </s> embedding as a dense vector for retrieval, while RankLLaMA concatenates query and document for scalar scoring. Together, they established new state-of-the-art results on MS MARCO DL (nDCG@10 = 77.9), eliminating heuristic pooling and proving that LLMs can encode long documents holistically. Unlike RankGPT's prompt-based reranking, fine-tuned LLaMA variants support efficient parallel inference, bridging the gap between academic benchmarks and practical deployment. These LLaMA-based models demonstrate the feasibility of globally encoding long documents and have achieved state-of-the-art performance on benchmarks such as MS MARCO. However, their context windows are still limited (2k–4k bytes), making them difficult to directly address the needs of long document IR. Furthermore, fine-tuning can lead to catastrophic forgetting, which degrades performance in short-text retrieval or cross-domain tasks, undermining their practicality.

\begin{table*}[ht]
\centering
\caption{Representative holistic approaches for long-document retrieval.}
\label{tab:holistic}
\resizebox{\textwidth}{!}{
\begin{tabular}{p{2.5cm} p{3.5cm} p{2.3cm} p{2.2cm} p{2.0cm} p{3.8cm}}
\toprule
\textbf{Method} & \textbf{Core Mechanism} & \textbf{Max Context Length} & \textbf{Engineering Complexity} & \textbf{IR Effectiveness} & \textbf{Key Limitations} \\
\midrule
\textbf{Naive Truncation (DPR, ANCE)} 
& Retain first 512 tokens of BERT; contrastive learning on truncated docs 
& 512 tokens 
& Low (standard PLM) 
& Effective for short passages; weak for long-doc IR 
& Severe information loss; positional bias; ignores discourse structure \\

\textbf{Longformer / BigBird} 
& Sparse/blockwise attention (sliding window + global tokens / random links) 
& 4k--16k tokens 
& Medium (custom CUDA kernels often required) 
& Strong on QA/summarization; moderate on IR 
& May miss cross-block evidence; implementation barrier \\

\textbf{Reformer / FlashAttention} 
& LSH attention / IO-aware exact attention 
& 8k--16k+ tokens 
& Medium--High (special kernels, memory tuning) 
& Efficient for long text modeling 
& Limited IR-specific optimization; harder training stability \\

\textbf{LONGEMBED} 
& Expand the context size of the embedding model 
& 32k tokens 
& Medium(training-free and training strategies) 
& Expand the length of the context. 
& Need to explore more context expansion methods based on training. \\

\textbf{QDS-Transformer} 
& IR-axiom-driven sparse pattern (locality, hierarchy, query matching) 
& 4k tokens 
& High (structured sparse kernels) 
& +3.25\% nDCG@10 on TREC DL’19 
& Task-specific design; limited generalizability \\

\textbf{Socialformer} 
& Small-world graph attention; dynamic token circles 
& 8k tokens 
& Medium (PyTorch only, no custom kernels) 
& Outperforms BigBird on TREC DL 
& Dynamic sampling may destabilize training; scaling beyond 10k unclear \\

\textbf{RankGPT} 
& GPT-4 as reranker; explicit listwise ranking via prompts 
& 8k--32k tokens (via sliding windows) 
& Low (prompting) 
& Strong reranking; distillable into small models 
& Expensive inference; window fragmentation; prompt sensitivity \\

\textbf{RepLLaMA / RankLLaMA} 
& Fine-tuned LLaMA for retrieval/reranking; holistic doc encoding 
& 2k--4k tokens (native LLaMA) 
& Medium (fine-tuning infra) 
& SOTA on MS MARCO DL (nDCG@10=77.9) 
& Context length limited; domain robustness concerns \\

\textbf{LongLoRA / LongLLMLingua / DuoAttention} 
& Sparse/grouped-query attention; prompt compression; head specialization 
& 8k--32k tokens 
& High (LLM training + compression) 
& Significant latency reduction; negligible quality loss 
& Faithfulness issues; still quadratic bottlenecks for very long docs \\

\bottomrule
\end{tabular}
}
\end{table*}

\subsubsection{Efficiency Challenges and Mitigation}
While holistic modeling offers strong representational capacity for long-document retrieval, its deployment in real-world systems is often constrained by efficiency bottlenecks. The primary challenges arise in several dimensions. First, the computational cost of Transformer-based architectures grows quadratically with sequence length, making it prohibitive to process ultra-long inputs or large document collections. Second, storage and indexing overheads increase significantly, as high-dimensional dense embeddings for long documents require substantial memory and incur latency . Third, inference throughput is limited by the high cost of large language model (LLM) rerankers, which are difficult to scale in latency-sensitive retrieval scenarios. Finally, multi-document interactions---a common requirement in information retrieval---further amplify the computational burden, since evidence must be aggregated across multiple long contexts.

To mitigate these issues, research has explored multiple complementary strategies. On the architectural side, sparse-attention mechanisms (e.g., Longformer, BigBird, QDS-Transformer) and hashing-based approximations (e.g., Reformer, Performer, FlashAttention) reduce the asymptotic complexity of self-attention, while hierarchical and graph-based models (e.g., Socialformer) capture global structure with improved scalability. Representation-level techniques such as multi-vector encoding (e.g., ColBERT, M3-Embedding), vector quantization, and knowledge distillation enable efficient storage and faster retrieval without severely degrading accuracy. At the retrieval pipeline level, multi-stage architectures remain dominant: a lightweight retriever conducts coarse filtering, followed by query-aware truncation, windowing, or LLM-based reranking. System-level optimizations, including batch inference, caching, and distributed ANN indexing, further enhance throughput in large-scale deployments.

Looking forward, efficiency will remain the central challenge for holistic long-document retrieval. Promising directions include adaptive retrieval strategies that dynamically allocate computation based on query complexity, tighter integration of retrieval and reranking models through distillation or adapters, and joint optimization across model, index, and hardware layers. Such approaches aim to bridge the gap between the holistic paradigm’s strong modeling capacity and the stringent efficiency requirements of real-world information retrieval systems.

\subsubsection{Limitations and Open Problems}
Holistic paradigms, while promising, remain constrained by efficiency, multimodality, and evaluation challenges. First, trade-offs between effectiveness and computational cost remain acute: million-token contexts (e.g., legal codes, technical manuals) remain impractical despite sparse attention and compression. Second, multimodal long documents: scientific papers with figures or news with embedded tables, remain largely unsupported; current VLMs (e.g., CLIP) lack robust alignment for 10k+ token multimodal contexts. Third, evaluation misalignment threatens progress: standard IR metrics (NDCG, Recall@k) fail to capture LLM-specific issues such as hallucinated evidence or failure to actually use context. Task-oriented metrics for faithfulness and attribution are urgently needed. Finally, robustness concerns including adversarial injections, pretraining contamination, and domain adaptation fragility underscore that LLMs cannot blindly replace IR pipelines. 

\noindent\textbf{Summary.}  
The holistic paradigm illustrates a clear trajectory: from truncation-limited PLMs, through sparse long-sequence Transformers, to end-to-end LLM rerankers. Each generation alleviates prior bottlenecks, but efficiency, faithfulness, and robustness remain fundamental challenges. For practical LDR, holistic approaches must be judiciously integrated with divide-and-conquer strategies, suggesting a hybrid future where block selection, efficient long-sequence encoding, and LLM reranking co-exist within the same pipeline.

\subsection{Divide-and-Conquer Paradigm for Long Documents (PLM \& LLM Era)}
\label{Divide-and-Conquer Paradigm methods}

In contrast to the holistic paradigm, the divide-and-conquer paradigm tackles the computational bottleneck caused by document length by segmenting long documents into smaller units, applying localized processing, and then aggregating the results. In this section, we will examine several key approaches within the divide-and-conquer paradigm, including pooling heuristics, hierarchical aggregation, and key block selection. While these methods are more computationally efficient, they face the challenge of effectively aggregating evidence dispersed across multiple blocks.

\begin{figure}[htbp]
    \centering
    \includegraphics[width=1.0\textwidth]{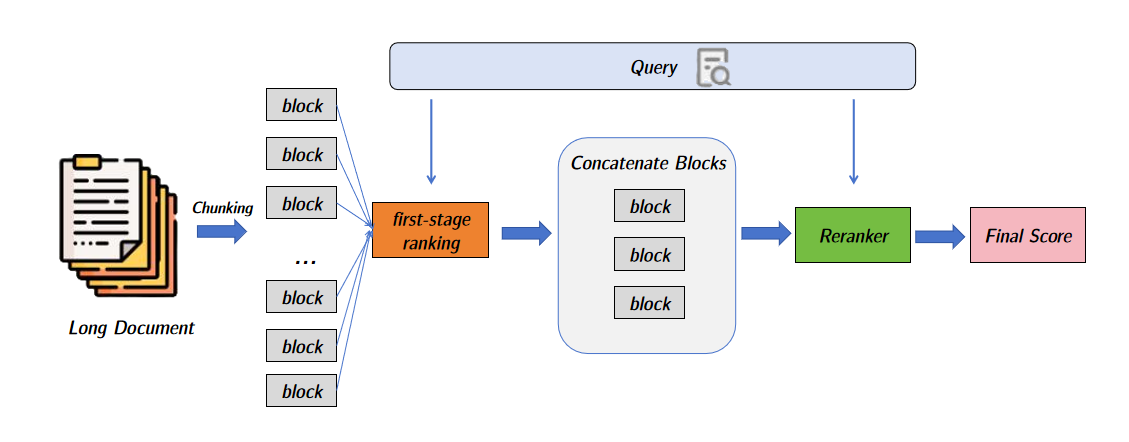} 
    \caption{A typical workflow for the key block selection approach within the divide-and-conquer paradigm, exemplified by models like KeyB. This approach concatenates the text of selected blocks before a final reranking.}
    \label{fig:divide_and_conquer}
\end{figure}

\subsubsection{Pooling-based Heuristics: BERT-MaxP and SumP}
The earliest adaptation simply applied BERT to fixed-length passages and aggregated passage scores.  
\citet{dai2019deeper} introduced \textbf{BERT-MaxP/SumP}, where each document $d$ is segmented into passages $\{p_1,\dots,p_n\}$; query–passage pairs $(q,p_i)$ are scored by a fine-tuned BERT cross-encoder, and document-level scores are aggregated:
\begin{align}
\mathrm{MaxP:}\; S(d,q) &= \max_i f_{\text{BERT}}(q,p_i), \\
\mathrm{SumP:}\; S(d,q) &= \sum_i f_{\text{BERT}}(q,p_i).
\end{align}
MaxP typically performs best, as it highlights the most relevant passage, but suffers when relevance is spread across multiple passages.  
Despite its simplicity, BERT-MaxP/SumP already outperformed strong lexical and neural baselines on Robust04 and ClueWeb09-B, especially for longer natural language queries.  
However, these heuristics inherit three key limitations: (i) distributed signals across passages may be lost; (ii) overemphasis on a single high-scoring passage risks missing complementary evidence; (iii) no mechanism for modeling document structure.

\subsubsection{Hierarchical Aggregation: PARADE, DRSCM, LTR-BERT, and MORES+}
To specifically address the loss of distributed signals and the overemphasis on a single high-scoring passage inherent in simple pooling, later works introduced explicit mechanisms to aggregate passage-level signals into document-level representations.

PARADE~\cite{li2023parade} encodes each $(q, p_i)$ pair using a cross-encoder PLM to obtain passage representations $h_i$, then aggregates them via multiple strategies, from simple pooling to learned networks. Among its learned strategies, PARADE-Attn uses a trainable vector to weigh passage importance, while the most powerful variant, PARADE-Transformer, employs a second Transformer encoder to model global dependencies between the passages. The final aggregated representation is then passed to a linear layer for scoring, allowing the model to form a holistic judgment based on evidence scattered across the text. The authors show that such learned aggregation methods consistently outperform simple heuristics, demonstrating that relevance in long documents often emerges from these cross-passage interactions.

DRSCM~\cite{wang2023novel} argues that relevance cannot be assessed locally alone, since topically divergent but locally relevant passages may dominate scores. To address this, it computes a segment correlation matrix offline, capturing the global centrality of each passage within the document. During online retrieval, passage scores are a linear combination of this global weight and the local query similarity, yielding robustness against topic drift. This architecture is highly efficient, as the computationally expensive correlation matrix is pre-computed, leaving only the lightweight combination for inference. These combined segment scores are then aggregated using a final pooling step (e.g., taking the maximum score) to produce the document-level ranking. DRSCM thus effectively bridges local query matching with the global semantic structure of the document.

LTR-BERT~\cite{wang2024efficient} decouples heavy offline document encoding from lightweight online query processing. Long documents are segmented and encoded offline, producing compressed passage embeddings. At query time, a short query is expanded and encoded online, then matched to stored embeddings using a parameter-free late interaction mechanism. This lightweight matching operates token-wise: for each query token, it finds the most similar document token within a passage, and the final score is computed by comparing the averaged vectors of the query and these selected best-match tokens. Notably, the model is trained efficiently on short-text pairs and uses a BERT-MaxP-style final aggregation, but achieves its speed by replacing the expensive cross-encoder with its parameter-free calculation. The result is a large efficiency gain: processing up to 333 times more documents per millisecond, while still outperforming cross-encoder baselines on MS MARCO document ranking.

MORES+~\cite{gao2022long} is a modular Transformer re-ranker that enables full query-to-document token interaction. Documents are chunked into segments, independently encoded by an encoder module, then jointly attended by a query-aware interaction module. Architecturally, it achieves this by using a BART encoder to process each chunk and a modified BART decoder for the joint query-to-all-chunk cross-attention. This design is highly efficient, maintaining linear complexity with respect to document length and supporting offline pre-computation of chunk representations to speed up inference. Unlike PARADE, MORES+ allows cross-attention across all document tokens, mitigating information loss from pooling. On Robust04 and MS MARCO, MORES+ outperformed PARADE and BERT-MaxP, establishing new state-of-the-art results at the time.

Hierarchical methods demonstrate that aggregating dispersed evidence is essential in long documents. PARADE and DRSCM emphasize learned and global aggregation, while LTR-BERT and MORES+ highlight architectural innovations for efficiency and fine-grained query–document interactions. Yet, their reliance on fixed segmentation risks fragmenting semantics, motivating selective and dynamic strategies.

\subsubsection{Key Passage/Block Selection: IDCM, KeyB, KeyB2, and DCS}
Another strategy is to first filter key content before deep re-ranking, reducing both cost and noise.

IDCM\cite{hofstatter2021intra} addresses the high query latency of re-ranking all passages in a long document with a powerful model. It employs a two-stage cascade where a lightweight "student" model first rapidly selects the top-$k$ most promising passages from within the document. Subsequently, a more powerful but computationally expensive "teacher" model re-scores only this small candidate set. Notably, the student model is trained to mimic the teacher's scoring behavior through a three-step knowledge distillation process, rather than being trained on ground-truth labels directly. This strategy allows IDCM to achieve effectiveness comparable to a full BERT-based evaluation while reducing median query latency by more than four times, crucially avoiding the need for expensive manual passage-level annotations.

KeyB~\cite{li2023power} analyzed that relevance signals are widely distributed but uneven in long documents. KeyB first selects key blocks using either a fast BM25 scorer or a more powerful learned BERT-based selector, then concatenates them for final ranking via a Transformer model like BERT or PARADE. Its most innovative variant, known as "BERT in BERT," cleverly reuses the same Transformer model to first score and select the key blocks and then to perform the final ranking on the concatenated result. While this learned selection scheme achieves state-of-the-art results on benchmarks like TREC DL, the simpler and much faster BM25-based selector offers a strong practical alternative by providing an excellent balance between performance and efficiency. This select-then-process strategy allows KeyB to consistently surpass sparse-attention models and pooling baselines.

KeyB2\cite{li2024keyb2} adapts the KeyB framework for the LLM era by employing a variety of selectors—including BM25, cross-encoders, and bi-encoders—to identify key blocks. These blocks are then processed by large LLM rerankers, such as LLaMA-3. In terms of effectiveness, this approach achieves state-of-the-art performance on the TREC DL benchmark, as measured by NDCG@10. The model also demonstrates significant efficiency gains, doubling the inference speed compared to full-document rerankers like RankLLaMA. This speedup is possible because the LLM only needs to process a small, highly relevant subset of the document's content. The success of KeyB2 demonstrates that even with powerful LLMs, strategic block selection remains crucial for reducing redundancy and focusing model capacity on the most salient evidence.

BReps~\cite{li2025enhanced} introduces a block-level representation framework for LDR. Documents are segmented into short, semantically coherent blocks, each encoded independently by a LLM. At query time, the query embedding is matched against all block embeddings, and the top-$k$ relevant block scores are aggregated to produce the document-level score. This design enables offline pre-computation of block representations and lightweight online scoring, while capturing fine-grained matching signals that are often lost in single-vector representations. Compared with dense (e.g., RepLLaMA) and late-interaction baselines, BReps achieves higher effectiveness, demonstrating consistent gains across multiple long document benchmarks.

DCS~\cite{sheng2025dynamic} addresses the incoherence of fixed-length segmentation through a two-stage process. First, for dynamic chunking, it identifies topic boundaries by calculating the semantic distance between adjacent sentences to create coherent segments. This is achieved by using Sentence-BERT embeddings and identifying split points where the cosine similarity between adjacent sentences is lowest, indicating a topic shift. Next, a lightweight classifier is trained to mimic the attention patterns of a more powerful teacher LLM via feature distillation. Specifically, the classifier learns to predict chunk relevance from features distilled from the teacher LLM's cross-attention matrix between the question and the context. This allows the classifier to efficiently select the most relevant chunks to feed into the final model. Experiments on long-context QA benchmarks show DCS robustly outperforms static chunking and maintains high accuracy even at 256k tokens.

Selection methods explicitly balance efficiency and effectiveness. IDCM cascades reduce latency; KeyB and KeyB2 show that intelligent block selection can even outperform full-document LLM rerankers; DCS demonstrates dynamic, semantically coherent segmentation. However, learned selectors may be costly to train, and the risk of missing critical evidence remains.

\subsubsection{Hybrid Cascades and Late Interaction: ICLI, Match-Ignition, and Longtriever}
Hybrid approaches combine multiple paradigms to jointly exploit local and global semantics.

ICLI\cite{li2022bert} model utilizes a single BERT architecture within a cascaded ranking process to balance efficiency and effectiveness. First, a fast pre-ranking stage uses the \texttt{[CLS]} token embedding of each passage to quickly identify a small set of top-$k$ candidates from the long document. Subsequently, a more precise but computationally expensive re-ranking stage, based on a ColBERT-style MaxSim operation, is applied only to this filtered set of passages. This entire two-stage process is trained end-to-end with multi-task learning, enabling the single model to master both tasks. The approach yields significant gains, improving NDCG@10 by 8\% over ColBERT\cite{khattab2020colbert} while achieving three times the inference speed of BERT-CAT.

Match-Ignition~\cite{pang2021match} is a hierarchical noise-filtering framework designed to mitigate the signal dilution problem common in long-document matching. It operates using a two-stage filtering cascade. First, a lightweight scorer prunes sentences with low cross-document similarity to remove noisy, unrelated content. Second, a word-level co-occurrence graph is constructed from the remaining text, and the PageRank algorithm is used to identify and filter out low-importance words. This aggressive distillation of the input allows a standard Transformer to match the documents' salient components without exceeding token limits, leading to strong performance on tasks like plagiarism detection and citation recommendation.

Longtriever~\cite{yang2023longtriever} addresses the limitations of traditional hierarchical models, where document blocks are often processed independently, leading to a loss of global context. Its architecture introduces a "tightly-coupled" interaction mechanism between local and global semantics. At each layer, an inter-block encoder first models the global context by attending over special [DOC] and [CLS] tokens from all blocks. This global context is then fed back into the intra-block encoders, allowing each block to be processed with an awareness of the full document's content. To combat annotation scarcity, Longtriever also introduces a novel pre-training task, Local Masked Autoencoder (LMAE), which learns to reconstruct tokens by fusing both global and local representations. This design leads to state-of-the-art performance on benchmarks like the MS MARCO Document Ranking and TREC Deep Learning tracks, outperforming other hierarchical and sparse-attention models. Furthermore, it achieves this effectiveness while maintaining a favorable efficiency profile, with a sub-quadratic complexity and inference latency comparable to other block-based methods.

Hybrid designs aim to unify efficiency and comprehensiveness. ICLI integrates dense retrieval with late interaction in a single architecture; Match-Ignition filters noise hierarchically; Longtriever couples local and global encoders. While effective, these models introduce additional architectural complexity and require careful efficiency–effectiveness trade-offs.

\subsubsection{Overall Assessment}
Divide-and-conquer remains the most mature and widely explored paradigm for adapting PLMs and LLMs to long documents. It offers practical compromises by decomposing input, but faces three persistent challenges: (i) semantic fragmentation from segmentation; (ii) risk of missing distributed signals in selection; (iii) computational inefficiency when scaling to millions of documents. 
Emerging LLM-based methods  suggest that intelligent selection, hierarchical modeling, and hybrid cascades can substantially mitigate these issues, yet further innovations in efficiency, robustness, and multimodal integration remain critical for real-world deployment.The persistent challenge of semantic fragmentation from fixed segmentation also motivated a different line of research that focuses not on the model architecture, but on the indexing structure itself, as discussed in the next section.

\subsection{Indexing-Structure-Oriented Paradigm}
\label{Indexing-Structure-Oriented Paradigm methods}
\begin{figure}[htbp]
    \centering
    \includegraphics[width=1.0\textwidth]{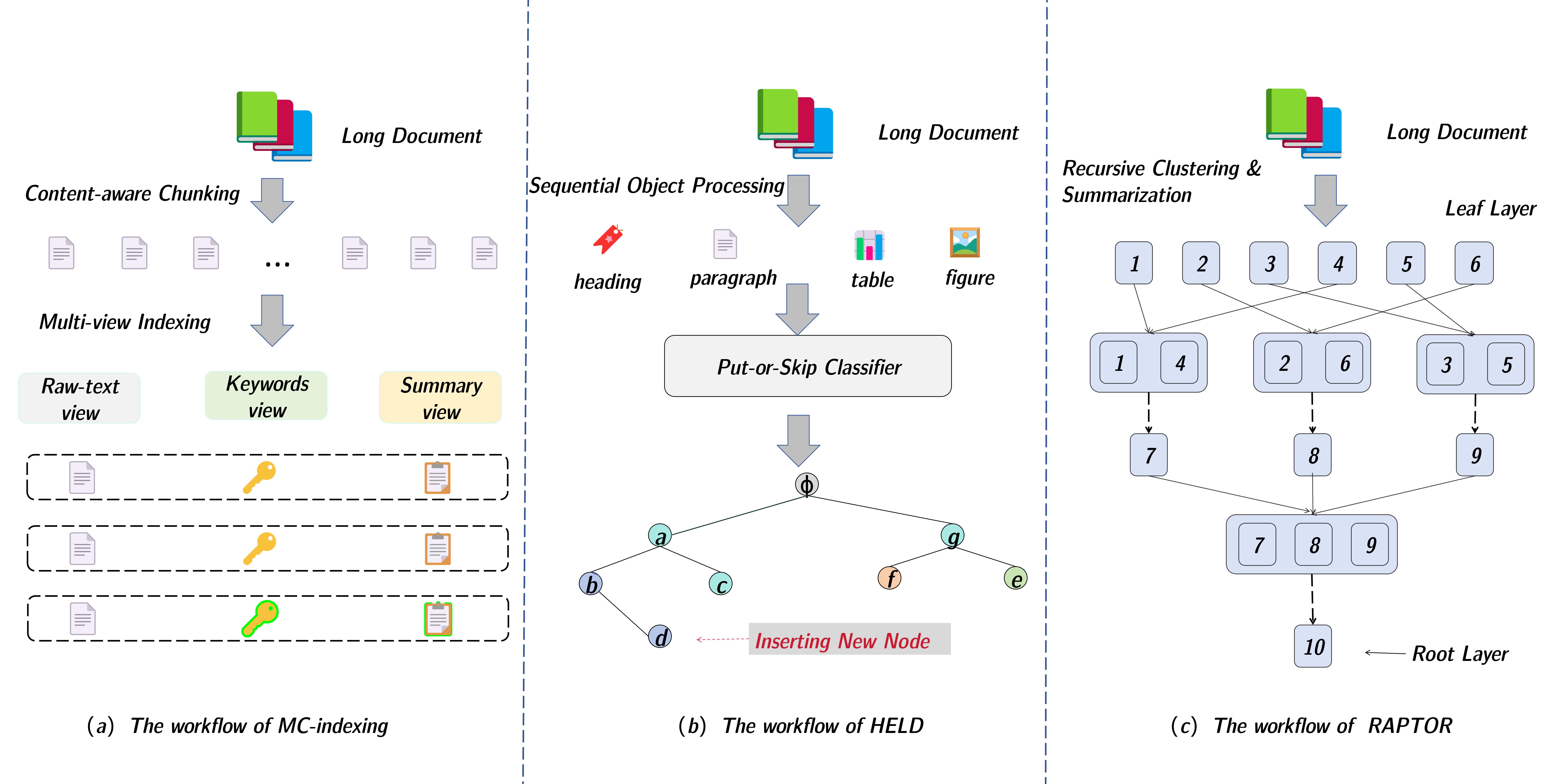}
    \caption{Conceptual overview of three indexing-structure-oriented paradigms: (a) MC-indexing, (b) HELD, and (c) RAPTOR.}
    \label{fig:indexing_paradigms_overview}
\end{figure}

While most long-document retrieval research focuses on model architectures 
, a complementary line of work emphasizes 
how documents are segmented and indexed. 
Naive fixed-length chunking often truncates relevant content or mixes unrelated 
information, limiting the effectiveness of even strong retrievers. 
Recent studies therefore redesign the indexing structure itself, providing 
orthogonal improvements that can seamlessly integrate with sparse or dense retrievers. 
This section reviews three representative paradigms, whose core workflows are conceptually illustrated in Figure~\ref{fig:indexing_paradigms_overview}.

MC-indexing~\cite{dong2024mc} improves retrieval by optimizing the indexing structure itself, rather than the retriever model. As illustrated in Figure~\ref{fig:indexing_paradigms_overview} (a), this process begins by segmenting a long document into semantically coherent units via "content-aware chunking." This initial step is critical, as it uses the document's header hierarchy to preserve semantic boundaries often broken by arbitrary fixed-length methods. Subsequently, each unit is indexed under three complementary views: the original raw text, an LLM-generated summary, and a set of extracted keywords. During retrieval, an off-the-shelf retriever scores each view independently before the results are aggregated. The method's unsupervised and plug-and-play design requires no retriever fine-tuning and can be seamlessly integrated with both sparse models like BM25 and various dense encoders. This leads to substantial recall improvements on long-document QA benchmarks, with gains of up to 40\% on WikiWeb2M.

HELD~\cite{cao2022extracting} model automatically extracts variable-depth logical hierarchies from long documents. Its workflow, depicted in Figure~\ref{fig:indexing_paradigms_overview} (b), innovatively mimics the human reading process. It sequentially processes document objects like paragraphs and tables, feeding them into a ``Put-or-Skip'' classifier. This classifier makes its decision by fusing both textual features and crucial visual cues, such as font size and indentation, from the document objects. This core mechanism then determines the correct attachment point for each object within a dynamically growing tree structure. To maintain efficiency and the document's original reading order, the model cleverly constrains this search to the "rightmost-branch" of the tree. A particularly effective two-step variant first constructs a tree from the document's headings before attaching content blocks. This approach not only achieves over 97\% accuracy on financial datasets but also significantly improves downstream passage retrieval, demonstrating the value of structural indexing over flat chunking.

RAPTOR\cite{sarthi2024raptor} organizes a document into a semantic tree through a process of recursive clustering and summarization. As visualized in Figure~\ref{fig:indexing_paradigms_overview}  (c), this bottom-up construction begins with initial text chunks (leaf nodes). These chunks are then recursively clustered based on semantic similarity and summarized by an LLM to form higher-level nodes. The diagram illustrates this iterative process, showing how multiple layers of summaries are built until a single root node is formed, effectively grouping semantically related information regardless of its original position in the text. For querying, RAPTOR's more robust "collapsed tree" strategy was found to be superior to a simple top-down traversal. This approach works by flattening all nodes from the tree (both original chunks and all levels of summaries) into a single pool. A standard dense retrieval is then performed across this entire pool, allowing the query to match information at any level of abstraction. By leveraging these multiple levels of abstraction, RAPTOR achieves notable gains on complex question-answering tasks. For instance, on the QuALITY dataset\cite{pang2021quality} , it boosts accuracy by over 8 percentage points compared to a DPR baseline, highlighting the potential of hierarchical, abstraction-aware indexing.

Collectively, these indexing-structure-oriented paradigms highlight three complementary strategies: MC-indexing enriches chunk representations through multiple views, HELD reconstructs a document's explicit logical hierarchy, and RAPTOR builds an implicit semantic hierarchy via recursive summarization. These methods underscore a crucial insight: retrieval effectiveness is not solely dependent on model architecture but is fundamentally constrained by how documents are segmented, represented, and indexed. Significant open challenges remain, particularly in handling heterogeneous document formats , integrating multimodal elements, and aligning these advanced indexing strategies with LLM-based retrievers for end-to-end optimization.

\subsection{Long-Query Retrieval: The Query-by-Document Task}
\label{subsec:qbd}
In specialized domains such as legal case retrieval, patent prior-art search, and scientific literature analysis, search tasks often move beyond short keywords, giving rise to the Query-by-Document  paradigm. In this setting, the query itself is a long, complex document used to find other semantically related documents. This "query-by-example" approach dramatically intensifies the core challenges of LDR. The problem of managing document length is squared---as both queries and candidates can span thousands of tokens---making efficiency and pre-computation paramount. Relevance signals also become more dispersed, demanding robust methods to aggregate conceptual similarity in the absence of strong lexical overlap. Consequently, conventional solutions are often inadequate: naive truncation risks information loss and positional bias, while exhaustive cross-encoder comparisons across all passages are computationally prohibitive.

To address these challenges, sentence-level bi-encoder models have been proposed, including one based on a \textbf{R}e-ranking with \textbf{P}roportional \textbf{R}elevance \textbf{S}core (RPRS)~\cite{askari2024retrieval}. Built on SBERT, RPRS operates by segmenting both the query $Q$ and candidate document $D$ into sentences.
Each sentence is then passed through an encoder $E(\cdot)$ to produce its vector embedding:
\begin{equation}
\mathbf{u}_i = E(q_i),\quad \mathbf{v}_j = E(d_j),\quad s_{ij}=\cos(\mathbf{u}_i,\mathbf{v}_j).
\end{equation}

Based on a similarity threshold $\tau$, it then computes two coverage proportions:
\begin{align}
\mathrm{QP}(Q\!\to\!D) &= \frac{1}{m}\sum_{i=1}^{m}\mathbb{I}\Bigl(\max_{j}\, s_{ij} \ge \tau\Bigr), \\
\mathrm{DP}(D\!\to\!Q) &= \frac{1}{n}\sum_{j=1}^{n}\mathbb{I}\Bigl(\max_{i}\, s_{ij} \ge \tau\Bigr),
\end{align}
where \(\mathrm{QP}\) measures the fraction of query sentences that find a close match in $D$, and \(\mathrm{DP}\) measures the fraction of document sentences covered by $Q$. The final Proportional Relevance Score is the product of these two components:
\begin{equation}
\label{eq:rprs}
S_{\text{RPRS}}(Q,D) = \mathrm{QP}(Q\!\to\!D) \times \mathrm{DP}(D\!\to\!Q).
\end{equation}
This sentence-level design is key to its effectiveness in QBD settings. It offers full-length coverage without truncation and is highly ANN-friendly, as document sentence embeddings can be pre-computed for scalable nearest-neighbor search. Furthermore, its reliance on coverage scoring requires low supervision, making it suitable for domains like legal search where labels are scarce. A frequency-aware variant, RPRS w/freq, extends this approach by incorporating BM25-style frequency saturation and length normalization to handle repeated matches more robustly~\cite{robertson1995okapi}.

In a modern retrieval pipeline, RPRS serves as a versatile and efficient re-ranker. By explicitly balancing query and document coverage, it avoids the pitfalls of methods that may over-emphasize a few salient passages while ignoring broad topical alignment. The source paper shows that while RPRS is robust, its effectiveness can be further improved by tuning its three parameters ($n, k_1, b$) on a small amount of labeled data. Future work  includes adapting the method for first-stage retrieval and exploring techniques to make it parameter-free by dynamically computing its parameters.

\section{Datasets and Empirical Benchmarks}
\label{sec:benchmarks}
This section summarizes representative datasets for evaluating long-document retrieval, together with evaluation protocols and benchmarking practices. We emphasize benchmarks where document length, dispersed evidence, or structure/layout are intrinsic to the task.

\subsection{Datasets}
\label{subsec:datasets}
\textbf{TREC DL Document Ranking Track (2019–2023).}
The document ranking task in the TREC Deep Learning (DL) track has been evaluated annually since 2019, with two subtasks: (i) full ranking from the entire corpus and (ii) top-100 re-ranking using organizer-provided candidates. Graded relevance labels (0–3) are used for testing, and NDCG@10 is the primary metric.
\begin{itemize}
    \item MS MARCO v1 (2019–2020): The 2019 and 2020 editions used the MS MARCO v1 document collection of approximately 3.21M web documents. For training, document labels were transferred from passage labels—a document was marked relevant if it contained a relevant passage. For evaluation, however, NIST provided independent human judgments at the document level (43 topics in 2019; 45 topics in 2020). Both full ranking and top-100 re-ranking subtasks were offered.
    \item MS MARCO v2 (2021–2023): From 2021, the track switched to the larger MS MARCO v2 corpus with about 11.9M documents. In 2021, document evaluation combined independent document judgments with labels propagated from passage annotations (57 topics). In 2022 and 2023, evaluation resources focused on building a reusable passage test set; the document task's labels were inferred from passage judgments (76 topics in 2022; 82 topics in 2023).
\end{itemize}

\textbf{TREC Robust04.}
The TREC Robust04\footnote{\url{https://trec.nist.gov/data/robust/04.guidelines.html}} dataset is a classic benchmark in the information retrieval field, widely used in research on long document retrieval tasks~\cite{Voorhees2004Robust}. This dataset originates from the TREC Robust Track 2004, with the core goal of evaluating the robustness of retrieval systems when facing difficult queries. The TREC Robust document collection is from TREC disks 4 and 5. It contains approximately 528,000 long news documents covering multiple news sources, 250 topics, and approximately 31,000 qrels. Relevance judgments in Robust04 are binary: documents are either marked as relevant (1) or non-relevant (0).

\textbf{GOV2 / Terabyte Track.}
Gov2\footnote{\url{https://ir.dcs.gla.ac.uk/test_collections/gov2-summary.htm}} is a large-scale web page corpus released by TREC in 2004 and is the core dataset of the TREC Terabyte Track (2004–2006)~\cite{Buttcher2006TrecTerabyte}. GOV2 contains documents resulting from a crawl of .gov
websites made in early 2004 and contains approximately 25 million web pages, with documents truncated to 256 kilobytes (KB), for an average document size of approximately 17.7 KB. It contains 150 queries with three level judgment: 0 (irrelevant), 1 (relevant) or 2 (very relevant).

\textbf{ClueWeb12.}
The ClueWeb12 dataset is designed to support research in information retrieval and related human language technologies. It contains 733,019,372 English web pages~\cite{clueweb12}. The dataset is divided into segments, each containing approximately 50 million documents. These documents are HTML web pages containing noisy content such as navigation, advertisements, and templates. Long, noisy pages make it suitable for testing robustness and dispersed relevance.

\textbf{MQ2007/2008.}
The MQ2007 and MQ2008 datasets are part of the LETOR~4.0 benchmark~\cite{qin2013introducing}, constructed from the GOV2 web collection of approximately 25 million documents and the TREC Million Query Tracks of 2007 and 2008. MQ2007 contains around 1,700 queries with roughly 65K judged query-document pairs, while MQ2008 includes about 800 queries with 14K pairs. Each query-document pair is represented by a 46-dimensional feature vector, incorporating various retrieval signals such as BM25 scores, language model features, and link-based features. Relevance annotations are graded on three levels (0, 1, 2), and the benchmark provides 5-fold cross-validation partitions for training, validation, and testing.
Although originally designed for learning-to-rank research, MQ2007/2008 are also cited in the context of long document retrieval because they are derived from full web pages in the GOV2 corpus. The relevance judgment of query-document labels enable  training and evaluation of long document IR models.

\textbf{MLDR.}
MLDR is a multilingual long document retrieval dataset based on Wikipedia, Wudao, and mC4, covering 13 typologically diverse languages~\cite{bge-m3}. It is constructed by extracting long articles from these datasets and randomly selecting paragraphs from them. GPT-3.5 is then used to generate questions based on these paragraphs. The generated questions and the extracted articles constitute the new text pairs in the dataset. It contains 41,434 training documents, 2,600 test documents, and 3,800 validation documents, with an average document length of 4,737.

\medskip
\noindent\textit{Domain corpora.} Collections such as PubMed (biomedical articles) and MIMIC-IV (clinical records)~\cite{johnson2020mimic} are rich sources for building long-context retrieval tasks; however, they are not standardized retrieval benchmarks by themselves and typically require task-specific query/label construction.

\subsection{Evaluation protocols and metrics}

This section reviews standard IR metrics that are also widely used for LDR. We first present their definitions, followed by a summary of potential challenges and considerations specific to LDR scenarios.

Let \(Q\) be the set of queries, \(L_q=\{d_1,\dots,d_N\}\) the ranked list for query \(q\), and \(\mathrm{rel}_i \in \{0,\dots,G\}\) the (graded) relevance of \(d_i\) at rank \(i\). Let \(\mathbb{I}[\cdot]\) be the indicator function and \(R_q=\{d:\mathrm{rel}(q,d)>0\}\) the set of relevant documents for \(q\).

\textbf{Precision@\(k\).}  
\begin{equation}
\mathrm{P@}k(q)=\frac{1}{k}\sum_{i=1}^{k}\mathbb{I}[\mathrm{rel}_i>0].
\end{equation}
Measures the proportion of relevant documents in the top-\(k\) results.

\textbf{Recall@\(k\).}  
\begin{equation}
\mathrm{Recall@}k(q)=\frac{\sum_{i=1}^{k}\mathbb{I}[\mathrm{rel}_i>0]}{|R_q|}.
\end{equation}
Evaluates coverage of relevant documents retrieved within the top-\(k\).

\textbf{Mean Reciprocal Rank (MRR).}  
\begin{equation}
\mathrm{MRR}=\frac{1}{|Q|}\sum_{q\in Q}\frac{1}{\mathrm{rank}_q},
\end{equation}
where \(\mathrm{rank}_q\) is the position of the first relevant document. Captures how early the first relevant document appears.

\textbf{Normalized Discounted Cumulative Gain (nDCG@\(k\)).}  
\begin{equation}
\mathrm{DCG@}k(q)=\sum_{i=1}^{k}\frac{2^{\mathrm{rel}_i}-1}{\log_2(i+1)},\qquad
\mathrm{nDCG@}k(q)=\frac{\mathrm{DCG@}k(q)}{\mathrm{IDCG@}k(q)}.
\end{equation}
Accounts for graded relevance and rank position through logarithmic discounting.

\textbf{Mean Average Precision (MAP).}  
\begin{equation}
\mathrm{AP}(q)=\frac{1}{|R_q|}\sum_{i=1}^{N}\mathrm{P@}i(q)\cdot \mathbb{I}[\mathrm{rel}_i>0],\qquad
\mathrm{MAP}=\frac{1}{|Q|}\sum_{q}\mathrm{AP}(q).
\end{equation}
Aggregates precision across recall levels to evaluate overall ranking quality.

Beyond accuracy, efficiency is also critical in LDR. Measures include end-to-end latency (per query), index size
and GPU usage.
These factors determine practical deployability and comparability across methods.

\medskip
\noindent\textbf{Discussion.}  
Traditional IR metrics remain the foundation of evaluation in LDR, but several characteristics of long documents complicate their interpretation. First, judgments are typically applied at the document level, which obscures finer distinctions such as “partially relevant” or “locally relevant” segments. Second, relevance evidence can be dispersed across different sections, while these metrics only consider whether the document as a whole is relevant. Third, coarse relevance scales are often insufficient to reflect nuanced utility in long documents. 
Future work may extend metrics to incorporate evidence localization and finer-grained relevance levels.

\section{Applications and Use Cases}
This chapter delves into the applications of LDR across various fields and the core challenges it faces. We begin by reviewing the progress of LDR technologies and, based on this foundation, analyze their specific applications in different domains, particularly highlighting their significance in law, academia, and life sciences. In addition, we explore the challenges that LDR encounters, such as efficiency bottlenecks, domain adaptability, and cross-modal retrieval issues.

\begin{figure}[htbp]
    \centering
    \includegraphics[width=0.8\textwidth]{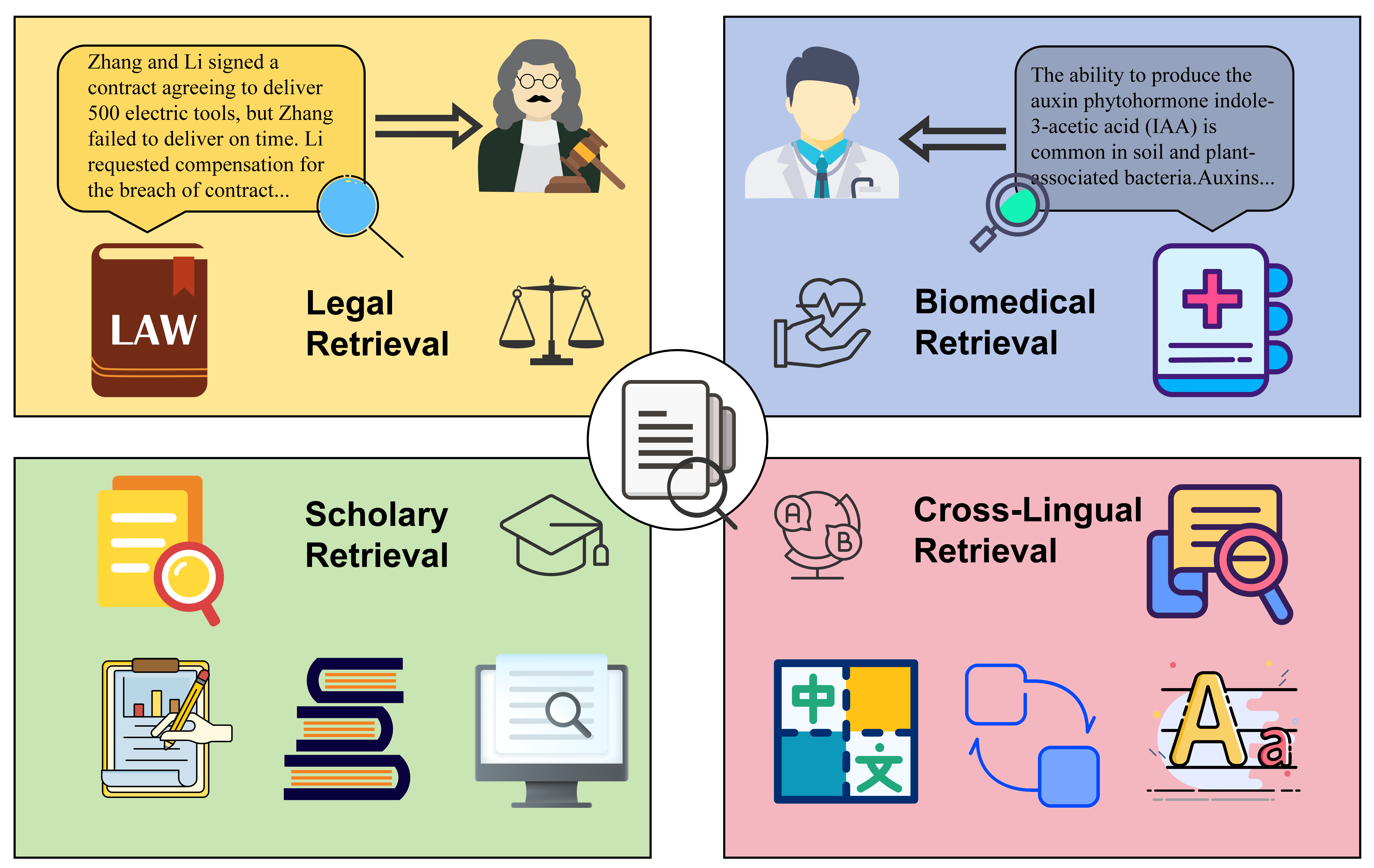} 
    \caption{The applications of Long-Document Retrieval} 
    \label{fig:long-document retrieval applications}
\end{figure}

Through Figure \ref{fig:long-document retrieval applications}, we present a schematic representation of LDR technologies applied across different fields, providing a clear framework for understanding how various tasks leverage LDR methods. Each domain in the figure corresponds to distinct technical requirements and solutions, such as long-document-based legal retrieval, academic paper retrieval, and cross-lingual retrieval. As LDR technologies continue to evolve, these applications demonstrate varying challenges and breakthroughs in practical implementation, especially when dealing with complex, structured documents and multimodal information retrieval.

\subsection{Legal Retrieval}
\label{subsubsec:legal-case-ldr}

Traditional keyword-based retrieval methods are rendered ineffective by the high specialisation, complex hierarchical structures, and extensive texts that characterise legal long-document retrieval. For instance, cross-version and cross-level mappings of provisions are frequently employed in judicial documents, legal provisions, and legal interpretations. The retrieval system must not only comprehend the rigors logic of legal language but also facilitate cross-document reasoning and query explainability. The system must distinguish between the consistency of legal principles and the similarity of case facts in case retrieval. Legal comparison necessitates the precise matching of changes in provisions across new and old versions. Comprehensive legal research necessitates the integration of case law, regulations, academic commentary, and judicial interpretations, resulting in a logically coherent chain of evidence. 

The research community has created a series of retrieval models that are specifically designed to address the characteristics of legal texts in order to overcome these challenges. Lawformer\cite{xiao2021lawformer}, which is constructed on the Longformer\cite{beltagy2020longformer} architecture, integrates global attention mechanisms with local sliding windows. This allows it to effectively manage judgement documents containing thousands of tokens and excel in legal question answering and case retrieval tasks .

Another model, Legal-BigBird\cite{dassi2021legal}, adapts the long-range Transformer model BigBird for legal document processing. It leverages BigBird's ability to handle long sequences with reduced computational costs, making it well-suited for the extensive length of legal texts. By further training BigBird on legal corpora, Legal-BigBird improves the representation of legal documents, enhancing its performance in tasks such as legal case retrieval.
Furthermore, LawGPT\cite{zhou2024lawgpt} is pretrained on a substantial corpus of legal literature, which includes legal judgements, regulations, and academic articles, in order to effectively convey the intricate logical relationships and profound semantics that are inherent in legal texts. LawGPT is capable of rapidly extracting pertinent information from lengthy legal documents during cross-document reasoning, thereby facilitating legal deduction and cross-document reasoning.

In conclusion, in order to effectively analyse intricate legal provisions, judicial rulings, and regulatory interpretations, legal lengthy document retrieval systems must possess a profound comprehension of the structure and semantics of legal documents. Secondly, the system must have the ability to integrate information from various sources in order to create a comprehensive legal evidence chain, which necessitates strong cross-document reasoning capabilities. Lastly, the system should improve its cross-lingual adaptability and explainability to guarantee its effective application in a variety of legal environments worldwide, thereby satisfying the requirements of judicial and legal research in various countries.

\subsection{Scholarly Long-Document Retrieval}
\label{subsec:scholarly-ldr}

Scholarly articles are long and multi-faceted (research questions, methods, datasets, results), so relevant evidence is dispersed across distant sections and easily diluted by length. Queries are often themselves long (e.g., paper-to-paper search), requiring decomposition into facet-specific intents and span-grounded verification to maintain provenance. Citation graphs additionally shape relatedness, making it necessary to reason over long-range, multi-hop links rather than purely local lexical overlap.

Citation- and concept-aware encoders provide high-recall seeds and stable geometry for long-corpus search: SciBERT supplies scientific-domain pretraining model for passage encodings~\cite{beltagy2019scibertpretrainedlanguagemodel}; SPECTER injects citation signals via triplet training (query paper, cited positive, uncited/hard-negative) to yield document embeddings aligned with scholarly relatedness, which is well-suited for document-level initial retrieval refined at passage level~\cite{cohan2020specterdocumentlevelrepresentationlearning}; SciNCL replaces discrete “cited/not” labels with neighborhood-contrastive sampling over a citation-graph embedding, improving global geometry for paper-to-paper search~\cite{ostendorff2022neighborhoodcontrastivelearningscientific}. 

Concept-indexed and aspect-aware retrieval narrows drift: SemRank builds a multi-granular concept index (topics + key phrases) per paper and uses LLM-guided concept selection at query time to match first at the concept layer before verifying at passage level~\cite{zhang2025scientificpaperretrievalllmguided}; PRISM decomposes a query paper into motivation/method/experiments subqueries, retrieves over a multi-vector corpus of 3k-token chunks, and fuses rankings (RRF) to operationalize “whole-document yet facet-specific” matching~\cite{park2025prismfinegrainedpapertopaperretrieval}.

To control cost on long texts, CORANK extracts zero-shot LLM features (categories/sections/keywords) to rerank large candidate pools, then applies full-text reranking only to the top-$M$ candidates, preserving LDR coverage while reducing token and latency budgets~\cite{tian2025llmbasedcompactrerankingdocument}. 

LLM-enhanced systems further tighten grounding and recall: PaperQA retrieves full texts, scores chunks with LLM-based relevance, and synthesizes answers with citations to curb hallucinations~\cite{la2023paperqaretrievalaugmentedgenerativeagent}; DocReLM trains retrievers/rerankers with LLM-generated pseudo-queries and traverses citation networks to gather supporting references~\cite{wei2024docrelmmasteringdocumentretrieval}; ensemble strategies such as LLM-KnowSimFuser fuse similarities from multiple LLM-enhanced embedder families with a LoRA-tuned academic encoder to stabilize ranking~\cite{dai2024advancingacademicknowledgeretrieval}. 

Multi-agent designs (e.g., PaSa, SPAR) modularize query understanding, citation-graph expansion, and reranking, harvesting long-range, multi-hop evidence while keeping verification passage-grounded~\cite{he2025pasallmagentcomprehensive,shi2025sparscholarpaperretrieval}.

In general, LDR for scholarly papers tames length by (i) abstracting queries to concepts/aspects to steer recall, (ii) leveraging citation graphs for long-range candidate proposals, and (iii) delaying full-text interaction to a narrow shortlist while enforcing span-level attribution to preserve provenance.

\subsection{Biomedical Literature Search}
\label{subsubsec:biomed-ldr}
High demands for multimodal fusion, strong knowledge interdependencies, and heterogeneous data structures characterise biomedical extended document retrieval, presenting unique challenges. The biomedical field's precise requirements are not satisfactorily addressed by conventional keyword-based retrieval methods. For instance, medical literature, electronic health records (EHR), and genetic reports require cross-disciplinary information mapping. This mapping must allow systems to interpret specialised terminology, support cross-modal reasoning, and result in traceability. Disease diagnosis support necessitates the integration of electronic health records (EHRs) with clinical guidelines, while clinical trial retrieval must align with trial designs, efficacy indicators, and patient characteristics. The integration of target literature, experimental data, and adverse reaction reports is necessary for drug repurposing .

To address the challenges of long clinical sequence processing, the research community has developed several models specifically designed for long-text retrieval. By optimizing the Transformer model, Clinical-Longformer and Clinical-BigBird\cite{li2022clinical} can process clinical texts with up to 4096 tokens. They adopt sparse attention mechanisms, which effectively reduce memory consumption and enable the processing of long texts without compromising performance. Meanwhile, they also excel at capturing long-range dependencies, making them perform exceptionally well in long-text retrieval tasks—particularly in clinical domain tasks such as question answering and document classification. These models can understand and retrieve complete clinical records in a single processing step, without splitting information into multiple segments, thus avoiding the issues of information loss or context fragmentation.

Additionally, BioClinical ModernBERT\cite{sounack2025bioclinical}, a domain-adapted encoder based on ModernBERT, also supports the needs of long-text retrieval and can handle longer contextual information. By integrating biomedical and clinical corpora, it not only significantly improves retrieval efficiency but also addresses the catastrophic forgetting problem commonly encountered in long-text processing. When processing large-scale patient health records, BioClinical ModernBERT can maintain the coherence and integrity of information, and it has demonstrated outstanding performance especially in tasks such as phenotype classification and clinical decision support.

Biomedical long-document retrieval relies on multimodal fusion and deep semantic understanding when processing massive and complex data. With the continuous advancement of model technologies, a range of advanced biomedical retrieval models have significantly enhanced their ability to process biomedical long-document by incorporating optimized attention mechanisms and deep learning approaches. The development of these technologies has not only promoted the accuracy and efficiency of literature retrieval but also provided crucial technical support for fields such as clinical decision-making and disease prediction. In the future, with the further expansion of data scale and continuous technological innovation, biomedical long-document retrieval will face more challenges and opportunities.

\subsection{Cross-Lingual Long-Text Retrieval}
\label{subsubsec:cross-lingual-ldr}
Cross-lingual information retrieval is particularly challenging when evidence resides in long documents such as encyclopedic articles or legal texts. Systems must simultaneously align semantics across languages and handle lengthy, structured content where relevant evidence is scattered.

A representative benchmark is XOR-QA\cite{asai2020xorqa}, which decouples the query language from the evidence language (e.g., a Japanese query with only English evidence). Since Wikipedia articles are long and multi-sectioned, models must not only retrieve passages across languages but also aggregate them for provenance-aware answers. Building on this benchmark, McCrolin\cite{limkonchotiwat-etal-2024-mccrolin} introduces a multi-consistency training framework with a teacher-student setup (frozen mUSE as teacher). It enforces cross-lingual semantic consistency and stable ranking via tailored loss functions, showing strong results on long-document retrieval across different encoders.

Beyond XOR-QA–based research, mGTE\cite{zhang2024mgtegeneralizedlongcontexttext} develops a long-context multilingual encoder (up to 8k tokens) with Rotary Position Embedding, unpadding, and two-stage pre-training, further enhanced by hybrid dense–sparse representations and a reranker to balance efficiency and accuracy. CROSS\cite{nezhad2025cross} targets ultra-long texts (up to 512k tokens) through a two-phase process: sentence-level retrieval with multilingual embeddings followed by selective reasoning with LLMs (e.g., GPT-4o-mini, Llama 3.2), effectively mitigating the “lost-in-the-middle” problem.

Although differing in architecture, context length, and optimization strategy, these approaches converge on the core pain points of cross-lingual long-text retrieval: semantic alignment, scattered evidence, and computational efficiency, while offering complementary solutions for cross-lingual long-context scenarios.

\subsection{Other Applications}

In addition to specialized fields such as legal, medical, and academic retrieval, long-document retrieval (LDR) has found widespread applications across various other domains. These applications have been driven by the explosive growth of digital content and the increasing complexity of information structures. As the scope of LDR continues to expand, it becomes crucial to explore how these retrieval systems adapt to broader and more complex domains. This section will examine these emerging applications, focusing on the unique challenges they present and the customized solutions developed to enhance retrieval performance.

\subsubsection{Web and News Retrieval}
With the explosive growth of digital content, long-document retrieval technology has shown great potential in the fields of web search and news retrieval. As information sources become increasingly diverse and content updates accelerate, traditional retrieval methods face unprecedented challenges. To address this issue, researchers have proposed multi-document retrieval tasks aimed at extracting relevant information sources from vast amounts of news articles to support efficient query execution. This task specifically emphasizes the high demands placed on retrieval systems due to the broadness of information and the diversity of sources in news reporting\cite{spangher2025novel}. The introduction of long-document retrieval technology enables efficient extraction of key data from lengthy news articles, blogs, and multimedia reports, and allows for real-time tracking of event developments, improving the speed and accuracy of information retrieval.

\subsubsection{Multimedia and Interactive Document Processing}
Modern long-document retrieval must increasingly contend with multimedia content, where meaning is conveyed through a complex interplay of text, layout, images, and tables. This has spurred the development of multimodal architectures and, critically, new evaluation frameworks to benchmark them. A key example is the MMDocIR benchmark~\cite{dong2025mmdocir}, which introduces page- and layout-level retrieval tasks specifically designed to assess performance on visually rich documents. The emergence of such dedicated benchmarks signifies a critical shift, pushing the field beyond text-centric models toward systems capable of genuine cross-modal reasoning and information fusion. M3DocRAG~\cite{cho2024m3docrag}  uses a multimodal retriever and MLM to find relevant documents and answer questions, allowing it to efficiently process single or multiple documents while preserving visual information.

\subsubsection{Enterprise Knowledge Management}
In the field of Enterprise Knowledge Management (EKM), as the volume of internal documents within companies continues to increase, traditional retrieval methods are increasingly inadequate to meet the need for efficient information retrieval. Research has shown that deep learning techniques demonstrate significant advantages in enterprise knowledge retrieval, effectively handling complex cross-domain knowledge retrieval tasks. With the generation and storage of large volumes of documents within organizations, long-document retrieval technology not only helps employees quickly find the most relevant information but also provides strong support in decision-making processes, thereby improving work efficiency. For example, the eSapiens system \cite{shi2025esapiens} combines a text-to-SQL planner with a hybrid Retrieval Augmented Generation (RAG) pipeline to support natural language access to both structured databases and unstructured text, further enhancing the accuracy and consistency of enterprise knowledge retrieval.

\section{Current Challenges and Future Directions}

Despite substantial progress, long-document retrieval (LDR) remains an open and rapidly evolving research frontier. Below we summarize critical challenges and outline promising research directions.

\subsection{Key Challenges}

\textbf{(1) Efficient Scaling.}
Even with long-context PLMs/LLMs (e.g., 32K–100K windows), end-to-end processing of full documents remains expensive. Sparse/kernelized attention, chunking, and compression reduce cost but can attenuate fine-grained matching signals that matter for retrieval and QA. A core challenge is to design architectures and indexing schemes that scale sub-linearly in document length while preserving retrieval fidelity. Systematic “scaling laws’’ for accuracy–cost trade-offs at extreme context lengths are still underexplored.

\textbf{(2) Relevance Localization and Multi-Granularity.}
Long documents interleave relevant and irrelevant content. Retrievers must localize small yet critical spans and synthesize document-level evidence. Existing block/segment selection (e.g., KeyB/KeyB2, hierarchical selectors, LLM-guided in-context selection) alleviates cost but can suffer from query drift, redundancy, or broken cross-block reasoning. Discourse- and structure-aware retrieval that respects document organization (sections, tables, figures, citations) remains limited.

\textbf{(3) Data Scarcity.}
High-quality document-level supervision is costly, making fully supervised learning difficult. Synthetic/weak supervision helps, but is often sentence-level and lacks document- or paragraph-granularity alignment. Leveraging cross-task signals (summarization, QA, citation linking) or implicit interaction signals is a promising route to richer supervision at document scale.

\textbf{(4) Domain and Language Adaptation.}
Specialized domains (legal, biomedical, scholarly) and multilingual LDR introduce domain-specific structures and terminology. Transfer from English-centric datasets remains challenging (e.g., MLDR-zh). Structured artifacts, e.g., clause numbering, formulas, tables, layouts are seldom integrated into retrieval models, limiting cross-domain generalization.

\textbf{(5) Interpretability and User Trust.}
Users need to understand why a long document is retrieved. Current neural/LLM models offer weak attribution: attention maps or post-hoc highlights may not reflect actual decision signals, and hallucination risks grow when models summarize long evidence. Span-level rationales, attribution-aware training, and user-centered evaluation protocols are urgently needed.

\subsection{Future Directions}

\textbf{Unified Retrieval–Reading Models.}
LLMs increasingly blur the retrieval–reasoning boundary by consuming long contexts. Jointly optimizing retrieval and reading, aided by context compression (e.g., LongLLMLingua~\cite{jiang2024longllmlingua}), selective pruning~\cite{li2023compressing}, and abstraction prompts, is promising. Open questions include when retrieval-free pipelines suffice, and how to characterize their cost–benefit at corpus scale.

\textbf{Advances in Long-Context Architectures.}
Progress in efficient attention (sparse/kernelized/chunked), improved positional schemes (e.g., RoPE variants~\cite{su2024roformer}), and external memory~\cite{wumemorizing} suggests hybrid designs that combine local/global attention with hierarchical representations. Structured or episodic memory that bridges symbolic storage (graphs, tables, layouts) and neural context modeling is a natural next step beyond 100K tokens.

\textbf{Retrieval-Enhanced LLMs.}
RAG for long documents remains under-engineered. Beyond plug-and-play retrieval, systems should support iterative retrieval-generation cycles~\cite{shao2023enhancing}, query refinement, and adaptive granularity (document/section/passage). Tighter integration of retrieval signals into LLM reasoning, e.g., retrieval-conditioned decoding, retrieval-aware training, may improve faithfulness and efficiency.

\textbf{Benchmarking and Evaluation.}
Conventional metrics (nDCG, Recall) overlook long-context issues such as attribution, multi-hop reasoning, and user effort. Many benchmarks emphasize short passages or moderate-length articles rather than truly long artifacts (patents, legal codes, books). Future resources should combine span-level and document-level annotations, include task-aware metrics (correctness vs.\ faithfulness), and report user-centric measures (e.g., time-to-answer, cognitive load) alongside latency and throughput.

\textbf{Cross-pollination and Multimodality.}
Insights from summarization, discourse parsing, and knowledge graphs can inspire structure-aware indexing. Many long documents are multimodal; aligning tables, figures, and text for joint indexing and retrieval remains open. Benchmarks such as VideoWebArena~\cite{jang2024videowebarena} expose large gaps to human performance; integrating layout modeling (e.g., LayoutLM-style features) and multimodal reasoning is a key avenue.

\subsection{Summary}
LDR sits at the intersection of IR efficiency, document understanding, and LLM reasoning. No single paradigm suffices: progress will hinge on hybrid solutions that combine sub-linear indexing, structure-aware modeling, and LLM-based reasoning with robustness, interpretability, and domain generalization. Equally important are realistic long-document benchmarks and user-centered evaluation to ensure LDR systems are trustworthy and deployable.

\section{Conclusion}

\label{sec:conclusion}

Long-Document Retrieval remains a fundamental yet unresolved problem in information retrieval, where the central challenge is to balance computational efficiency with the ability to accurately locate sparse but critical signals within lengthy contexts. Through this survey, we traced the evolution of the field from classical lexical approaches to neural architectures, and further to the recent integration of large language models. 

We synthesized existing methods into four complementary paradigms: Holistic modeling of entire documents, Divide-and-Conquer strategies that segment and re-aggregate evidence, Indexing-Structure innovations that exploit document organization, and specialized approaches for long-query retrieval. This taxonomy highlights the shift from static pipelines toward more dynamic, agentic retrieval systems empowered by LLMs. Across diverse domains such as legal case analysis, biomedical literature retrieval, and scientific paper search, these technologies are already demonstrating their capacity to mitigate information overload. Yet, our review underscores that no single paradigm offers a universal solution. Instead, the optimal choice is inherently application-dependent, demanding a careful balance of trade-offs between computational efficiency, inference latency, and retrieval effectiveness, guided by the specific nature of the documents and user information needs.

Looking ahead, we argue that the most promising progress will come from hybrid architectures: systems that fuse the efficiency of classical IR indexing, the semantic depth of neural encoders, and the reasoning capabilities of LLMs. Equally important is the development of next-generation evaluation frameworks that move beyond coarse-grained metrics to capture attribution, faithfulness, and user-centered utility in long-document settings. As these hybrid systems mature, they promise to transform how we interact with large-scale information, moving beyond simple document retrieval towards a future of deep evidence synthesis and automated knowledge discovery at scale.


\bibliographystyle{ACM-Reference-Format}
\bibliography{sample-base}

\end{document}

%% file: new_taxonomy.tex
\definecolor{fill_0}{RGB}{251, 245, 251}
\definecolor{fill_1}{RGB}{246, 249, 255}
\definecolor{fill_2}{RGB}{244, 249, 241}

\definecolor{draw_0}{RGB}{168, 80, 168}
\definecolor{draw_1}{RGB}{54, 110, 210}
\definecolor{draw_2}{RGB}{112, 173, 71}

\definecolor{fill_leaf}{RGB}{248, 248, 248}
\definecolor{draw-leaf}{RGB}{135, 135, 135}

\tikzstyle{my-box}=[
    rectangle,
    draw=draw-leaf,
    rounded corners,
    text opacity=1,
    minimum height=1.5em,
    minimum width=5em,
    inner sep=2pt,
    align=center,
    fill opacity=.5,
    line width=0.8pt,
]
\tikzset{
leaf/.style={
my-box,
minimum height=1.5em,
fill=fill_leaf,
text=black,
align=left,
font=\footnotesize,
inner xsep=2pt,
inner ysep=4pt,
line width=1pt
}
}

\begin{figure*}[!t]
\centering
\begin{adjustbox}{width=\textwidth}
\begin{forest}
forked edges,
for tree={
    grow=east,
    reversed=true,
    anchor=base west,
    parent anchor=east,
    child anchor=west,
    base=center,
    font=\large,
    rectangle,
    draw=draw-leaf,
    rounded corners,
    align=left,
    text centered,
    minimum width=5em,
    edge+={darkgray, line width=1pt},
    s sep=3pt,
    inner xsep=2pt,
    inner ysep=3pt,
    line width=1.2pt,
    ver/.style={rotate=90, child anchor=north, parent anchor=south, anchor=center},
},
where level=0{text width=10.5em,fill=fill_0,draw=draw_0,font=\large,}{},
where level=1{text width=11.5em,fill=fill_1,draw=draw_1,font=\normalsize,}{},
where level=2{text width=14em,fill=fill_2,draw=draw_2,font=\normalsize,}{},
where level=3{font=\footnotesize,}{},
[
    Long-Document \\ Retrieval, minimum height=2.8em
    [
        Pre-PLM Era \\ (Sec 3)
        [
            Lexical Methods \\ (Sec 3.1)
            [
                { TF-IDF~\cite{JONES1973619}, BM25~\cite{robertson1995okapi}, Heuristic Chunking~\cite{10.1007/978-1-4471-2099-5_31}}, leaf, text width=42em
            ]
        ]
        [
            Early Neural IR \\ (Sec 3.2)
            [
                { DSSM~\cite{10.1145/2505515.2505665}, DRMM~\cite{10.1145/2983323.2983769}, MatchPyramid~\cite{10.5555/3016100.3016292}, PACRR~\cite{hui2017pacrr}, DeepRank~\cite{pang2017deeprank}}, leaf, text width=42em
            ]
        ]
    ]
    [
        Core Paradigms \\ PLM \& LLM Era \\ (Sec 4)
        [
            The Holistic Paradigm \\ (Sec 4.1)
            [
                {\textbf{\textit{Naive Truncation}}: DPR~\cite{karpukhin2020dense}, ANCE~\cite{xiong2020approximate}}, leaf, text width=42em
            ]
            [
                {\textbf{\textit{Long-Sequence Transformers}}: Longformer~\cite{beltagy2020longformer}, BigBird~\cite{zaheer2020big}, Reformer~\cite{kitaevreformer}, FlashAttention~\cite{dao2022flashattention}, \\ LongNet~\cite{ding2023longnet}, LONGEMBED~\cite{zhu2024longembedextendingembeddingmodels}, M3-Embedding~\cite{bge-m3}, QDS-Transformer~\cite{jiang2020long}, Socialformer~\cite{zhou2022socialformer}}, leaf, text width=42em
            ]
            [
                {\textbf{\textit{LLMs as Rerankers}}: RankGPT~\cite{sun2023chatgpt}, RepLLaMA/RankLLaMA~\cite{ma2024fine}}, leaf, text width=42em
            ]
            [
                {\textbf{\textit{Efficiency Mitigation}}: LongLoRA~\cite{chen2023longlora}, LongLLMLingua~\cite{jiang2024longllmlingua}, DuoAttention~\cite{xiaoduoattention}}, leaf, text width=42em
            ]
        ]
        [
            The Divide-and-Conquer \\ Paradigm (Sec 4.2)
            [
                {\textbf{\textit{Pooling-based Heuristics}}: BERT-MaxP/SumP~\cite{dai2019deeper}}, leaf, text width=42em
            ]
            [
                {\textbf{\textit{Hierarchical Aggregation}}: PARADE~\cite{li2023parade}, DRSCM~\cite{wang2023novel}, LTR-BERT~\cite{wang2024efficient}, MORES+~\cite{gao2022long}}, leaf, text width=42em
            ]
            [
                {\textbf{\textit{Key Block Selection}}: IDCM~\cite{hofstatter2021intra}, KeyB~\cite{li2023power}, KeyB2~\cite{li2024keyb2}, 
                BReps~\cite{li2025enhanced},
                DCS~\cite{sheng2025dynamic}}, leaf, text width=42em
            ]
            [
                {\textbf{\textit{Hybrid \& Late Interaction}}: ICLI~\cite{li2022bert}, Match-Ignition~\cite{pang2021match}, Longtriever~\cite{yang2023longtriever}}, leaf, text width=42em
            ]
        ]
        [
            Indexing-Structure- \\ Oriented Paradigm \\ (Sec 4.3)
            [
                { MC-indexing~\cite{dong2024mc}, HELD~\cite{cao2022extracting}, RAPTOR~\cite{sarthi2024raptor}}, leaf, text width=42em
            ]
        ]
        [
            Long-Query Retrieval \\ (Sec 4.4)
            [
                { RPRS~\cite{askari2024retrieval}}, leaf, text width=42em
            ]
        ]
    ]
    [
        Evaluation \\ (Sec 5)
        [
            Datasets and Benchmarks \\ (Sec 5.1)
            [
                {TREC Robust04~\cite{Voorhees2004Robust}, Gov2/Terabyte Track~\cite{Buttcher2006TrecTerabyte}, ClueWeb12~\cite{clueweb12},  TREC Deep Learning~\cite{craswell2020overview}, MS MARCO~\cite{bajaj2016msmarco},\\ MLDR ~\cite{bge-m3}, LONGEMBED~\cite{zhu2024longembedextendingembeddingmodels}, Multi-CPR (Chinese)~\cite{Long2022MultiCPRAM}}, leaf, text width=42em
            ]
        ]
        [
            Standard Metrics \\ (Sec 5.2)
            [
                { nDCG@k, MAP, Recall@k, Precision@k, MRR,ERR}, leaf, text width=42em
            ]
        ]
    ]
    [
        Applications \\ (Sec 6)
        [
            Legal Retrieval \\ (Sec 6.1)
            [
                {Lawformer~\cite{xiao2021lawformer}, LegalBERT~\cite{chalkidis2020legal}, LawGPT~\cite{zhou2024lawgpt}, ChatLaw~\cite{cui2023chatlaw}, DISC-LawLLM~\cite{yue2023disc}}, leaf, text width=42em
            ]
        ]
        [
            Scholarly Paper Retrieval \\ (Sec 6.2)
            [
                {SciBERT~\cite{beltagy2019scibert}, SPECTER~\cite{cohan2020specterdocumentlevelrepresentationlearning}, SciNCL~\cite{ostendorff2022neighborhoodcontrastivelearningscientific}, PRISM~\cite{park2025prismfinegrainedpapertopaperretrieval},  CORANK~\cite{tian2025llmbasedcompactrerankingdocument}, PaperQA~\cite{la2023paperqaretrievalaugmentedgenerativeagent}, SemRank~\cite{zhang2025scientificpaperretrievalllmguided}, \\DocReLM~\cite{wei2024docrelmmasteringdocumentretrieval},  LLM-KnowSimFuser~\cite{dai2024advancingacademicknowledgeretrieval}, PaSa~\cite{he2025pasallmagentcomprehensive}, SPAR~\cite{shi2025sparscholarpaperretrieval}}, leaf, text width=42em
            ]
        ]
        [
            Biomedical Retrieval \\ (Sec 6.3)
            [
                {PubMedBERT~\cite{gu2021domain}, Clinical-Longformer, Clinical-BigBird~\cite{li2022clinical}, BioClinical ModernBERT~\cite{sounack2025bioclinical}}, leaf, text width=42em
            ]
        ]
        [
            Cross-Lingual Retrieval \\ (Sec 6.4)
            [
                {mBERT~\cite{devlin2018bert}, LaBSE~\cite{feng2020labse}, XLM-R, McCrolin~\cite{limkonchotiwat-etal-2024-mccrolin}, mGTE~\cite{zhang2024mgtegeneralizedlongcontexttext}, CROSS~\cite{nezhad2025cross}}, leaf, text width=42em
            ]
        ]
    ]
]
\end{forest}
\end{adjustbox}
\caption{A structured taxonomy of Long-Document Retrieval, categorizing existing research across eras, core paradigms, applications, and evaluation methods.}
\vspace{-4ex}
\label{fig:taxonomy_ldr}
\end{figure*}